\documentclass[final, 1p, times, authoryear]{elsarticle}
\usepackage{amssymb}
\usepackage{amsthm}
\usepackage[colorlinks=true, allcolors=blue]{hyperref}     
\usepackage{bm}
\usepackage[ruled,vlined]{algorithm2e}
\usepackage{algorithmic}
\usepackage{amsmath}
\usepackage{array}
\usepackage{multirow}
\usepackage{xcolor}   
\usepackage[subpreambles=true]{standalone}

\begin{document}


%% \author{Chen Wang}
%% \ead{chenw3@illinois.edu}
%% \corref[label1]{Chen Wang}

%% \fntext[label1]{Corresponding author. Email address: chenw3@illinois.edu}
\begin{frontmatter}
\title{Hierarchical Bayesian Modeling for Time-Dependent Inverse Uncertainty Quantification}


%% \author[label1]{Chen Wang }

\author{Chen Wang\corref{cor1}\fnref{label1}}
\cortext[cor1]{Corresponding author. Email address: chenw3@illinois.edu}

\affiliation[label1]{organization={Department of Nuclear, Plasma and Radiological Engineering, University of Illinois Urbana Champaign}}

\begin{abstract}

This paper introduces a novel hierarchical Bayesian model specifically designed to address challenges in Inverse Uncertainty Quantification (IUQ) for time-dependent problems in nuclear Thermal Hydraulics (TH) systems. The unique characteristics of time-dependent data, such as high dimensionality and correlation in model outputs requires special attention in the IUQ process. By integrating Gaussian Processes (GP) with Principal Component Analysis (PCA), we efficiently construct surrogate models that effectively handle the complexity of dynamic TH systems. Additionally, we incorporate Neural Network (NN) models for time series regression, enhancing the computational accuracy and facilitating derivative calculations for efficient posterior sampling using the Hamiltonian Monte Carlo Method - No U-Turn Sampler (NUTS).

We demonstrate the effectiveness of this hierarchical Bayesian approach using the transient experiments in the PSBT benchmark. Our results show improved estimates of Physical Model Parameters’ posterior distributions and a reduced tendency for over-fitting, compared to conventional single-level Bayesian models. This approach offers a promising framework for extending IUQ to more complex, time-dependent problems.

\end{abstract}

\begin{keyword}
Hierarchical Bayesian Model \sep Time-Dependent Modeling \sep Inverse Uncertainty Quantification \sep Thermal Hydraulics \sep Markov Chain Monte Carlo \sep Surrogate Models 


\end{keyword}
\end{frontmatter}

%%\linenumbers

\section{Introduction}
\label{introduction}

In recent years, the field of nuclear Thermal Hydraulics (TH) has increasingly relied on computational simulations to predict the behavior of nuclear reactor systems under various conditions. These simulations are crucial for ensuring the safety and efficiency of nuclear reactors. However, they are inherently uncertain due to limitations in accurately modeling complex physical phenomena. This uncertainty becomes particularly pronounced in time-dependent simulations, where the dynamics of the system introduce additional layers of complexity. Inverse Uncertainty Quantification (IUQ) has emerged as a pivotal methodology to address these uncertainties, specifically by quantifying the uncertainties associated with Physical Model Parameters (PMPs) in these systems.

The final report of the BEMUSE project (\cite{perez2011uncertainty}) underscores the necessity of input uncertainty quantification as a pivotal element in probabilistic uncertainty analysis within Best Estimate plus Uncertainty (BEPU) frameworks. The prevalent approach in BEPU methodologies involves the transmission of uncertainties from inputs to model predictions, making the determination and validation of uncertainty ranges for each parameter vital. Traditionally, these ranges have been subjectively set based on expert opinions, highlighting the need for a more empirical and structured method. This issue was a key focus of the international PREMIUM project (\cite{skorek2019quantification}), though it concluded without a unified agreement on the procedures and guidelines for input uncertainty quantification (UQ). The absence of a consensus led to the initiation of the OECD/NEA SAPIUM project (Systematic APproach for Input Uncertainty quantification Methodology) (\cite{baccou2019development} \cite{baccou2020sapium}), which sought to establish a methodical approach to input UQ in nuclear thermal-hydraulics (TH) codes. Following this, in 2022, the OECD/NEA launched the ATRIUM project (Application Tests for Realization of Inverse Uncertainty quantification and validation Methodologies in thermal-hydraulics) \cite{ghione2023applying}, with the main objective being the practical implementation of IUQ exercises to demonstrate the efficacy of the SAPIUM approach.

The landscape of IUQ has been considerably shaped by the progressive advancements in machine learning (ML) and artificial intelligence (AI). These technologies have significantly improved the precision and dependability of simulation models in various sectors, tackling a multitude of intricate challenges~\cite{wang2023scientific}. Their successful deployment spans several disciplines, including healthcare (\cite{dong2021influenza,dong2021semi,chen2019claims}), agriculture (\cite{wu2022optimizing}), transportation (\cite{ma2022application,meng2022comparative,li2023exploring}), clinical research (\cite{xue2021use,xue2022perioperative}), signal processing (\cite{hu2022dan,li2023nst, liu2021measuring}), structural health monitoring (\cite{liu2023intelligent}), reliability engineering (\cite{chen2020optimal, wang2024optimal, chen2017multi, chen2020some}), industrial engineering (\cite{chen2018data, li2023applying, chen2018data, chen2023recontab, wu2024switchtab}), and artificial intelligence (\cite{liu2019dapred, liu2023stationary, liu2024cliqueparcel}). The adaptability and success of ML/AI approaches in these areas provide compelling evidence of their potential and impart crucial learnings for advancing IUQ initiatives within the nuclear power industry.

Traditional IUQ approaches, predominantly using single-level Bayesian models, have provided significant insights into steady-state TH systems. However, they exhibit limitations when applied to applications with various experimental conditions and large datasets. Key challenges include handling the high variability of PMPs under dynamic experimental conditions and avoiding over-fitting due to unknown model discrepancies or outliers. Recent advancements in hierarchical Bayesian models ~\cite{wang2023inverse,wang2023scalable}  have shown promise in addressing these issues in steady-state nuclear TH systems, but their applicability for time-dependent problems remains underexplored.

This study aims to fill this gap by introducing a novel hierarchical Bayesian model tailored for time-dependent IUQ in nuclear TH systems. The main objectives are twofold: firstly, to develop a model capable of handling the unique challenges of time-dependent data, such as high dimensionality and correlation; and secondly, to demonstrate the hierarchical model's efficacy in providing more accurate and reliable estimates of PMPs' posterior distributions, while reducing the tendency for over-fitting. The study leverages advanced statistical techniques, including Gaussian Processes (GP) integrated with Principal Component Analysis (PCA), and Neural Network (NN) models for time series data regression. Hamiltonian Monte Carlo Method - No U-Turn Sampler (NUTS) are used  for efficient posterior sampling. The method's performance is evaluated using the transient experiments in the PSBT benchmark, setting a new precedent in the application of hierarchical Bayesian methods to time-dependent IUQ in nuclear TH systems.

The rest of the paper is organized as follows. Section \ref{sec2} will give an overview of Hierarchical Bayesian model for IUQ. Section \ref{sec3} will introduce the overview of the PSBT benchmark and TRACE modeling. Section \ref{sec4} will introduce the surrogate models for time-dependent problems, and then in Section \ref{sec5}, the hierarchical framework is applied to a case study for TRACE physical model parameters using the PSBT benchmark data. Section \ref{sec6} will be the summary.

\section{Hierarchical Bayesian Framework for Inverse Uncertainty Quantification}
\label{sec2}

\begin{figure}[!htbp]
    \centering
    \includegraphics[width = \textwidth]{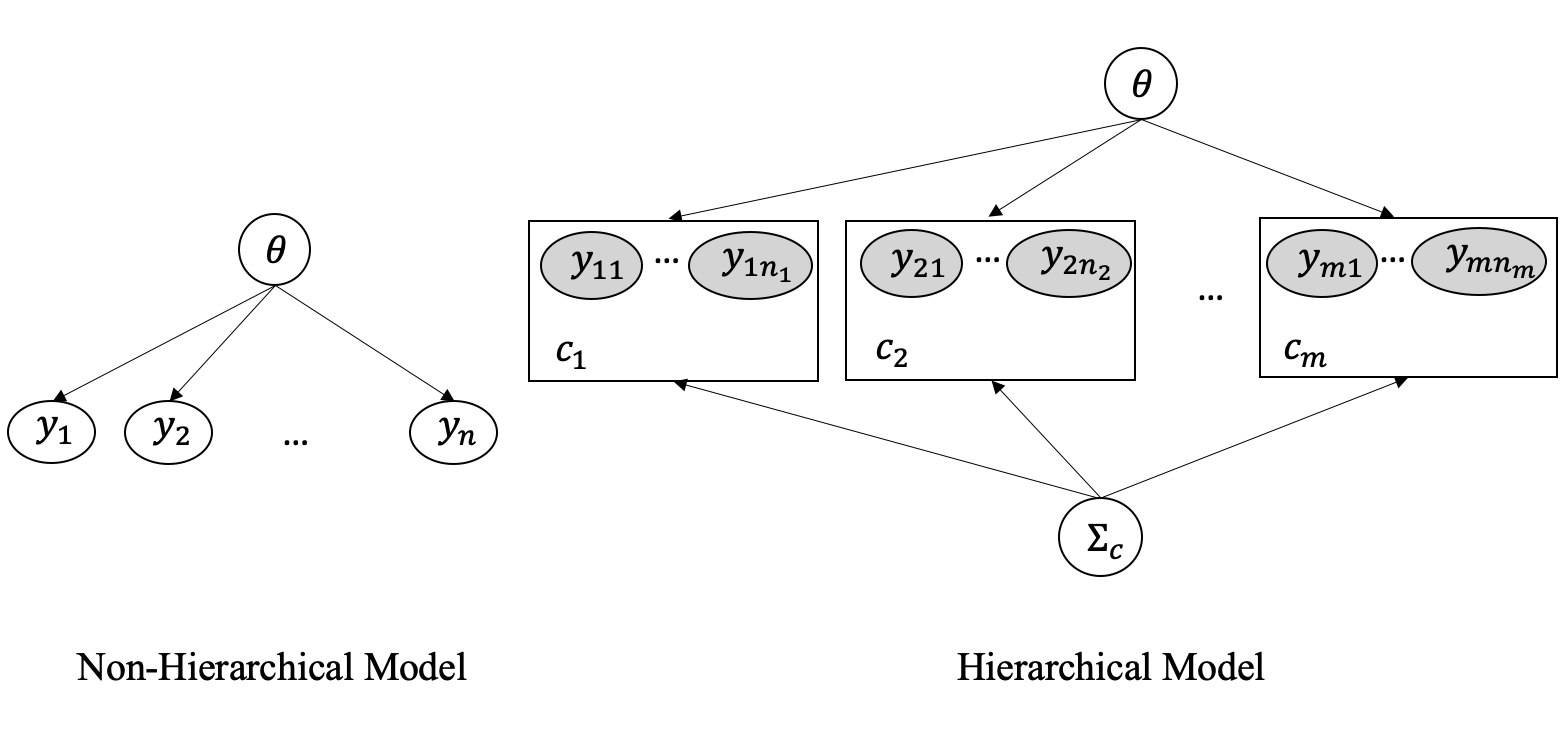}
    \caption{Non-hierarchical and hierarchical models }
    \label{fig:hb1}
\end{figure}

Hierarchical model is also called a multi-level model or mixed-effect model. A simple illustration of the structure of the hierarchical model is shown in Figure \ref{fig:hb1}. Observations $y$ are from different clusters determined by (1) cluster parameter $c_m$ which might be different among clusters, and (2) parameter $\bm \theta$ shared across clusters. If the probability distribution of $c_m$ can be parameterized by $\bm \Sigma_c$, $\bm \Sigma_c$ can be seen as the cluster-specific parameter, or per-group parameter, to be estimated (\cite{wang2020hierarchical}).

Consider a simple example for Figure \ref{fig:hb1}, where the observations are normally distributed. In this case, the parameters we need to estimate are the mean and the variance. Suppose they have the same variance $\sigma^2$ for all clusters but different means $\mu_i$ for each cluster. So the $\sigma^2$ is a shared parameter ($\theta$ in Figure \ref{fig:hb1}) and $\mu_i$ is per-group parameter ($c_i$ in Figure \ref{fig:hb1}). Now we would like to specify the distribution over the cluster-specific $\mu_i$. We can assume the distribution also normal, or other distributions depending on the actual scenario. If normal distribution is assumed, two parameters (global mean $\mu$ and global standard deviation $\sigma_y$, correspond to the $\Sigma_c$ in Figure \ref{fig:hb1}) would be required. So we can describe parameter $\mu_i$ and observations $y_{ij}$ as:
$$
\mu_i \sim N(\mu, \sigma_y^2)
$$
$$
y_{ij} \sim N(\mu_i, \sigma^2 )
$$

\begin{figure}[!htbp]
    \centering
    \includegraphics[width = 0.8\textwidth]{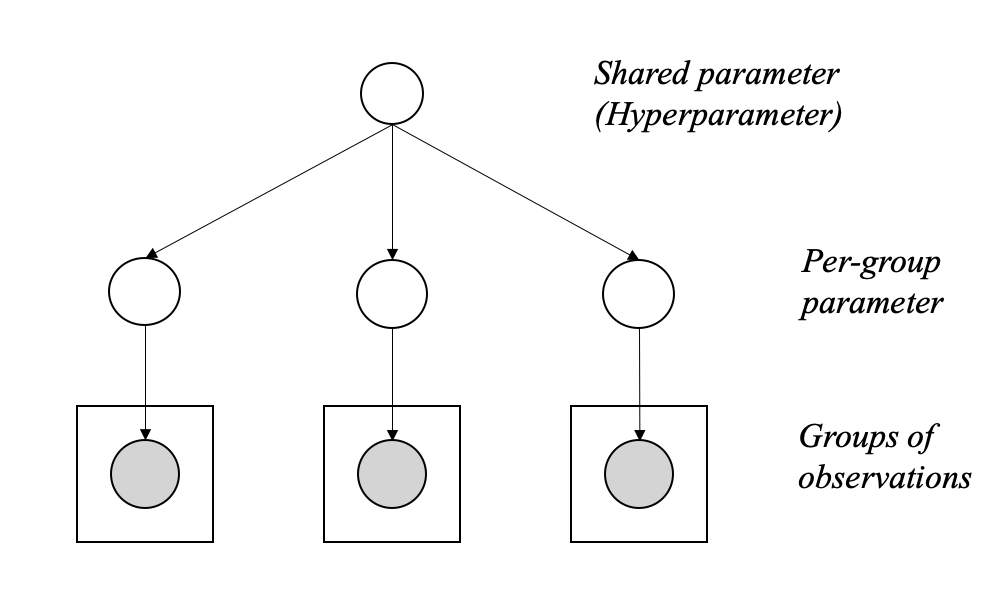}
    \caption{Illustration of a classical hierarchical model structure}
    \label{fig:hb2}
\end{figure}

Now we can re-arrange the structure of the above hierarchical model to better illustrate the process of the above two equations: the per-group parameters are generated from shared parameters, and observations are generated from per-group parameters, as shown in Figure \ref{fig:hb2}. 

In summary, the hierarchical model comes in handy when we know that some cluster-level attributes exist and the structure of the data can be reasonably modeled or assumed. When data follows a hierarchical structure, a simple non-hierarchical model would be inappropriate because (1) a model with few parameters cannot fit the dataset accurately, and (2) a model with many parameters will ``overfit'' the dataset. In contrast, a hierarchical model can have enough parameters to fit the data, while has a population distribution to model the dependence of the parameters. Thus ``overfitting'' can be avoided because the parameters are ``constrained`` by the population distribution so they cannot fit the data exactly.

\begin{figure}[!htbp]
    \centering
    \includegraphics[width = 0.4\textwidth]{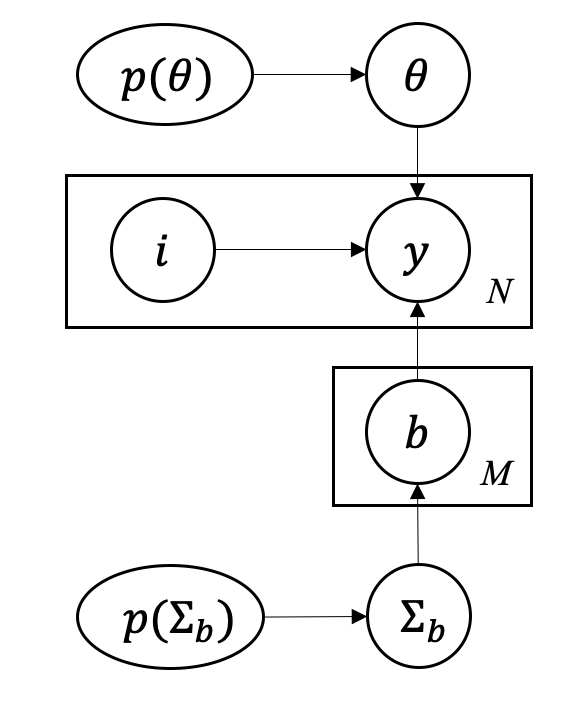}
    \caption{Graphical representation of a hierarchical model}
    \label{fig:hb3}
\end{figure}

The hierarchical model structure in Figure \ref{fig:hb1} can be plotted into a more succinct and formal graphical representation, as shown in Figure \ref{fig:hb3}. In the figure, $p(\theta)$ is the prior distribution of $\theta$, $i$ is the cluster index, $y$ is the observation, and there are $N$ observations in total. the per-group parameter $b_i$ is governed by its distribution $\Sigma_b$, which has prior distribution $p(\Sigma_b)$.\\

It should be noted that in this model, our primary goal is to learn about $\theta$ and $\Sigma_b$, not the individual cluster-specific parameters $b_i$, but $b_i$ also has to be estimated in order to estimate $\Sigma_b$. So we need to marginalize over the $b_i$ parameters in after we have obtained the joint posterior distribution. There are generally two approaches to estimate the posterior, we can either use MLE for point estimation or use Bayesian inference to get full posterior distributions for these parameters. 

In the MLE method, the likelihood of $\theta$ and $\Sigma_b$, marginal over $b$, can be expressed as:
\begin{equation}
    L(\bm \Sigma_b, \bm \theta; \bm y) = \int_b P(\bm y \mid \bm \theta, \bm b, i)P(\bm b|\bm \Sigma_b) \,db
\end{equation}

If full posterior distribution is desired, we can incorporate the prior distribution and use Bayes' rule. The prior distribution can be decomposed into the following form:
\begin{equation}
    P(\bm \Sigma_b, \bm \theta, b_i) = P(b_i \mid \bm \Sigma_b) \cdot P(\bm \Sigma_b, \bm \theta)
\end{equation}
which is based on the fact that $\Sigma_b$ influences observations only through $b_i$. So the marginalized joint posterior distribution can be expressed as:
\begin{equation}
    P(\bm \Sigma_b, \bm \theta \mid \bm y) \propto \int_b P(\bm y \mid \bm \theta, \bm b, i)P(\bm b \mid \bm \Sigma_b) \cdot P(\bm \theta)P(\bm \Sigma_b) \,d b
\end{equation}

Hierarchical models provide a flexible framework for modeling the complex interactions but come with higher computational requirements in inference. The parameter space to be estimated can be high-dimensional: we need to estimate cluster-specific parameter $b_i$, which may contain up to hundreds. Traditional random-walk based MCMC algorithms such as MH or advanced MH algorithms tailored for relatively high dimensional problems still suffer from serious convergence issues because that the random behavior of the proposal function in very inefficient in high-dimension domains. HMC and NUTS would be useful in such high-dimensional conditions.
\section{Transient Experiments in PSBT Benchmark}
\label{sec3}
OECD/NRC benchmark based on NUPEC PWR subchannel and bundle tests (PSBT) is designed for validation purposesof void distribution in subchannel and PWR bundle and pre-diction of departure form nucleate boiling Uncertainty Quantification for Steady-State PSBT Benchmark using Surrogate Models. 
The void distribution benchmark in PSBT includes transient bundle benchmark, which can be applied to system TH codes to assess their capabilities of predicting the void generation during transients (\cite{specifications2010oecd}). The experimental data in these transients include X-ray densitometer measurements of void fraction (chordal averaged) at three axial elevations. The averaging is over the four central subchannels. Data is collected for four transient scenarios: Power Increase (PI), Flow Reduction (FR), depressurization (DP), Temperature Increase (TI), and at three different assembly types 5, 6, and 7. All 5, 6, and 7 assemblies are $5 \times 5$ rob bundles while 5 and 6 have typical cells and 7 has thimble cells. They also have different axial and radial power distributions. The PSBT benchmark has been used extensively in many IUQ related applications (\cite{borowiec2017uncertainty,wang2018ans})

\begin{figure}[!htbp]
    \centering
    \includegraphics[width = 0.8\textwidth]{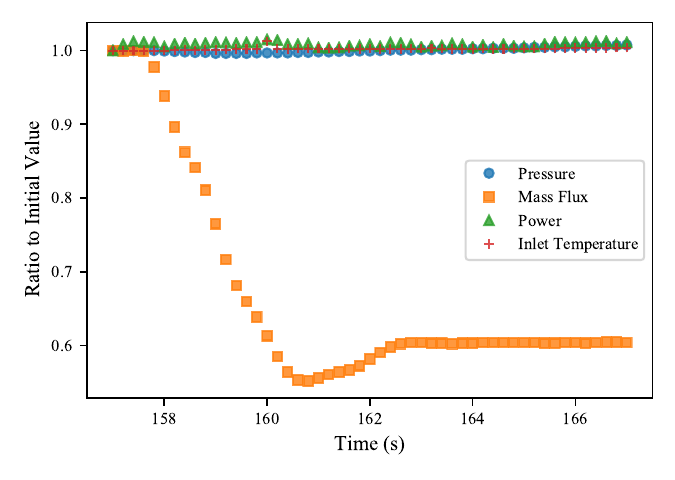}
    \caption{Boundary conditions during the 7T-flow reduction transient}
    \label{fig:bc-ts}
\end{figure}

In this work, we managed to build TRACE models for power increase, flow reduction, and temperature increase transients for all three assemblies. Figure \ref{fig:bc-ts} shows the variation of boundary conditions in 7T-flow reduction transient: during the flow reduction transient, flow rate was reduced and other boundary conditions were maintained at the same level.

\begin{figure}[!htbp]
    \centering
    \includegraphics[width = 0.7\textwidth]{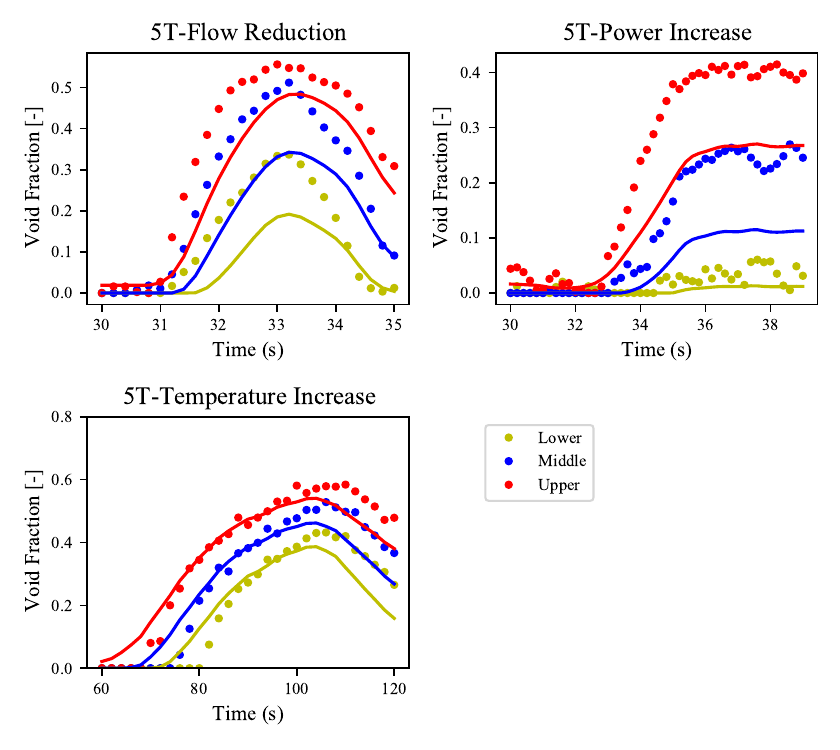}
    \caption{Transients simulation in Assembly 5. (Dots: Experiment, Lines: Simulation)}
    \label{fig:ts_5vali}
\end{figure}

\begin{figure}[!htbp]
    \centering
    \includegraphics[width = 0.7\textwidth]{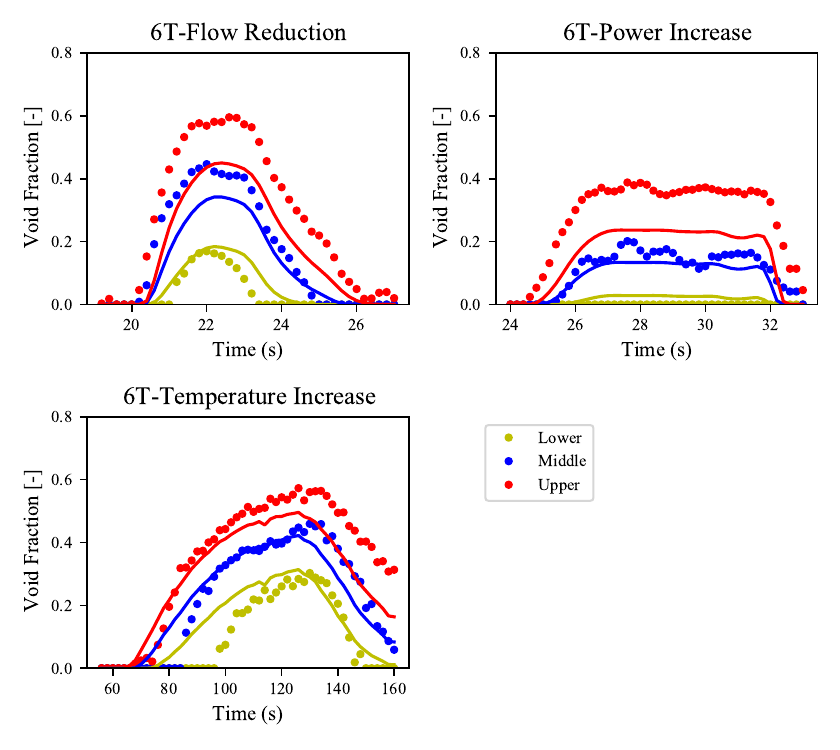}
    \caption{Transients simulation in Assembly 6. (Dots: Experiment, Lines: Simulation)}
    \label{fig:ts_6vali}
\end{figure}

\begin{figure}[!htbp]
    \centering
    \includegraphics[width = 0.7\textwidth]{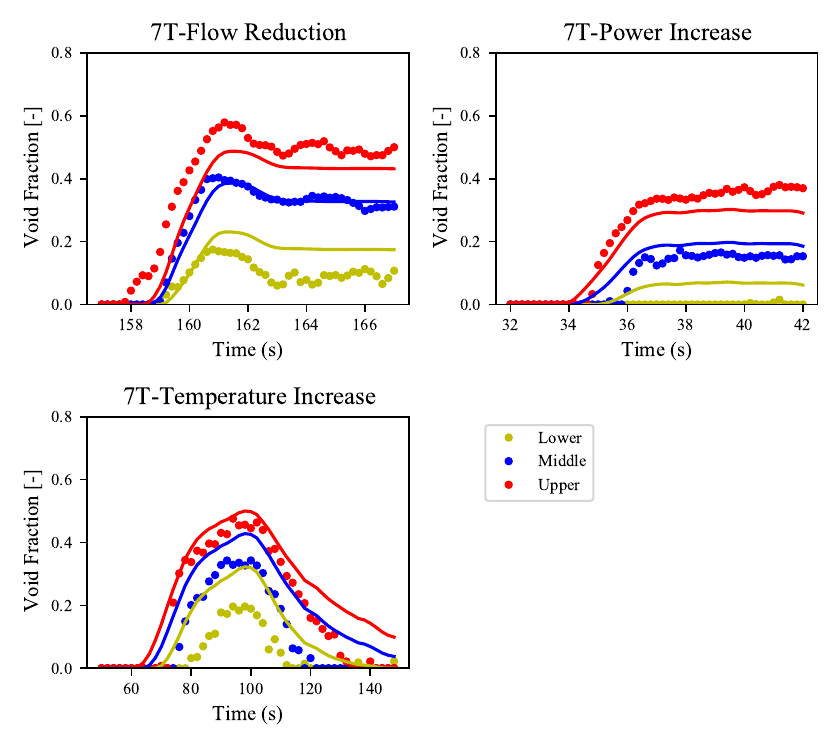}
    \caption{Transients simulation in Assembly 7. (Dots: Experiment, Lines: Simulation)}
    \label{fig:ts_7vali}
\end{figure}

Figure \ref{fig:ts_5vali}, \ref{fig:ts_6vali}, and \ref{fig:ts_7vali} show comparisons between TRACE predicted void fraction and measured void fraction for the three assemblies, respectively. Dots represent experiment measurements and lines are TRACE simulations. We can see that in some cases, TRACE performs well and have good agreement with experimental data, while there are also obvious model discrepancies in some transients.


Since the transient and the steady-state experiments share same assemblies and similar boundary conditions, it is reasonable to use the same physical model parameters as we have selected in previous similar studies (\cite{wang2017sensitivity, wang2019gaussian, wangsurrogate, wangnureth3}). The four parameters are listed in Table \ref{tab:6parameter_2}.

\begin{table}[!htbp]
    \centering
    \caption{List of 4 selected physical model parameters in TRACE}
    \begin{tabular}[c]{m{5em}  c}
    \hline
    \textbf{Parameter Number}     &  \textbf{Definition} \\
    \hline
    P1008     & Single phase liquid to wall heat transfer coefficient \\
    P1012     & Subcooled boiling heat transfer coefficient   \\
    P1022     & Wall drag coefficient \\
    P1028     & Interfacial drag (bubbly/slug Rod Bundle-Bestion) coefficient\\
    \hline
    \end{tabular}
    \label{tab:6parameter_2}
\end{table}

\section{Surrogate Model for Time-Dependent Thermal-Hydraulics Systems}
\label{sec4}

Surrogate model, metamodel or emulator plays a vital role in the Bayesian calibration of computer models when the computer models are computationally expensive, such as nuclear reactor simulations.
The surrogate models are built from limited number of runs of the full model at specifically selected values of the input parameters and a learning algorithm. The Design of Experiment (DoE) or computer experiments design~\cite{santner2013design}, in the context of surrogate models, aims to select a structured set of tests of the computer model, such that a good tradeoff between the accuracy of the surrogate model and the number of tests can be achieved. A good coverage of the input space will ensure the accuracy of the emulator while too many samples will cause extra computational burden, which is against the purpose of using the surrogate models. Latin Hypercube Sampling (LHS) is a widely-used method to generate samples from a multidimensional distribution. It has good space-coverage property and will be used in this work. 

Many types of surrogate models have been utilized in the Bayesian calibration process, such as the Polynomial Chaos Expansion (PCE) and Stochastic Collocation (SC) used by ~\cite{wu2017inverse} ~\cite{wu2017inverse2}, the Multivariate Adaptive Regression Splines (MARS) proposed by~\cite{stripling2013calibration}, the Function Factorization with Gaussian Process (FFGP) developed by~\cite{yurko2015demonstration}, and the deep Neural Networks by~\cite{liu2018data}. There are many more applications that cannot be included here. Different types of surrogate models have different characteristics, so they may fit in different engineering scenarios. Some factors to consider in choosing a proper one include the dimension of inputs, non-linearity of the input-output relationship, and time-dependent or steady-state problem, etc. 

\subsection{Gaussian Processes}
\label{sec:gp}
Gaussian Processes have been widely used as surrogate models since the seminal work of Bayesian calibration by~\cite{kennedy2001bayesian} utilized GP modeling. In this section, we will firstly go through the theoretical backgrounds of GP (\cite{wang2019gaussian, wangnureth1}), and look into several characteristics of GP.

A Gaussian process can be used to describe a distribution over functions. It is normally defined as \textit{a collection of random variables, any finite number of which have a joint Gaussian distribution} (\cite{rasmussen2004gaussian}). A GP is completely defined by its mean function $m(\bm{x})$ and covariance function $k(\bm{x},\bm{x}')$, and can be written as:
\begin{equation}
    f(\bm{x}) \sim \mathcal{GP} (m(\bm{x}), k(\bm{x}, \bm{x}'))
\end{equation}
where the random variables in the process represent the value of the function $f(\bm{x})$ at location $\bm{x}$. For notation simplicity we will treat the mean function as zero at this time. Later we will see this is not a drastic limitation because the posterior mean is not confined to be zero, and the prior mean function does not have major influence in interpolation.

Given training inputs and outputs $\left \{ (\bm{x_i},f_i)\mid i=1,...n \right \}$, we want to predict the test outputs $\mathbf{f_*}$ in test points $X_*$. The prediction outputs (including training outputs $\mathbf{f}$ and test outputs $\mathbf{f_*}$) are jointly distributed as a multivariate normal distribution:
\begin{equation}
    \begin{bmatrix} 
    \mathbf{f}\\\mathbf{f}_*\end{bmatrix}\sim N \Bigg( 0, \begin{bmatrix} K(X, X), K(X, X_*)\\K(X_*, X), K(X_*, X_*)\end{bmatrix} \Bigg)
\end{equation}
where $K(X, X_*)$ denotes the matrix of the covariances evaluated at all pairs of training and test points, and similar definitions apply to $K(X, X)$, $K(X_*, X_*)$, and $K(X_*, X)$. We are interested in the conditional probability of $ \{\mathbf{f}_*|X_*, X,\mathbf{f}\}$, which, fortunately, is simple to obtain due to properties of multivariate normal distributions.
\begin{equation}
\label{eqa:posteriorgp}
\begin{split}
    \mathbf{f}_* \mid X_*, X,\mathbf{f} \sim N \Big( & K(X_*, X)K(X, X)^{-1}\mathbf{f}, \\
    & K(X_*, X_*)-K(X_*, X)K(X, X)^{-1}K(X, X_*) \Big)
\end{split}
\end{equation}

Now we can evaluate the function values $\mathbf{f}_*$ according to Equation (\ref{eqa:posteriorgp}) by calculating the corresponding mean and the covariance matrix. The posterior means are used as prediction values. The covariance matrix represent the level of confidence in the prediction, and it will enter the covariance term $\bm{\Sigma}$ in Equation \ref{eqa:sigma_total}, so that the covariance term in the likelihood will include the error information from both the experiment and the emulator: $\bm{\Sigma} = \bm{\Sigma_{\text{exp}}}+ \bm{\Sigma_{\text{emu}}}$. This treatment is important because inaccurate predictions by the emulator can be seen as assigned a much smaller weight in evaluating the likelihood in Equation \ref{eqa:sigma_total}. This unique characteristic of GP makes it a preferable choice of an surrogate model in the Bayesian Calibration process.

The covariance function is a crucial ingredient in a GP predictor as it encodes the information of nearness or similarity between data points. This information is crucial because in regression tasks we normally assume closer input points are more likely to have similar target values, and thus the training points which are nearer to a test point should have more influence on the prediction at that point. The choice of the covariance function should depend on the assumption of the function which we wish to learn. In most of the cases and in this work it is defined as a function of the Euclidean distance between two points, with several hyperparameters. For example, power-exponential covariance function:
\begin{equation}
    \text{cov}(f(\bm{x}_p),f(\bm{x}_q)) = k(\bm{x}_p,\bm{x}_q)= \sigma^2\cdot \exp (-\frac{1}{2l^2}\mid \bm{x}_p-\bm{x}_q \mid ^r)
\end{equation}

The hyperparameters $\sigma,l,$ and $r$ define the proprieties of the covariance function such as magnitude, shape, and smoothness. They should be properly estimated to obtain the best possible prediction performance. 

There is also Gaussian Kernel, where $r$ is simply changed to 2:
\begin{equation}
    k(\bm{x}_p,\bm{x}_q)= \sigma^2\cdot \exp (-\frac{1}{2l^2}\mid \bm{x}_p-\bm{x}_q \mid ^2)
\end{equation}

The Mat\'ern class of kernel is given by:

\begin{equation}
    k(\bm{x}_p,\bm{x}_q)= \frac{1-\nu}{\Gamma(\nu)} (\frac{\sqrt{2\nu} \mid \bm{x}_p-\bm{x}_q \mid }{l})^{\nu} K_{\nu}(\frac{\sqrt{2\nu} \mid \bm{x}_p-\bm{x}_q \mid }{l})
\end{equation}
with positive parameters $\nu$ and $l$, and $K_\nu$ is a modified Bessel function (\cite{rasmussen2004gaussian}). $\nu = 3/2$ and $\nu = 5/2$ are popular choices. Different kernels may fit in different scenarios.

\begin{figure}[!h]
    \centering
    \includegraphics[width =0.8 \textwidth]{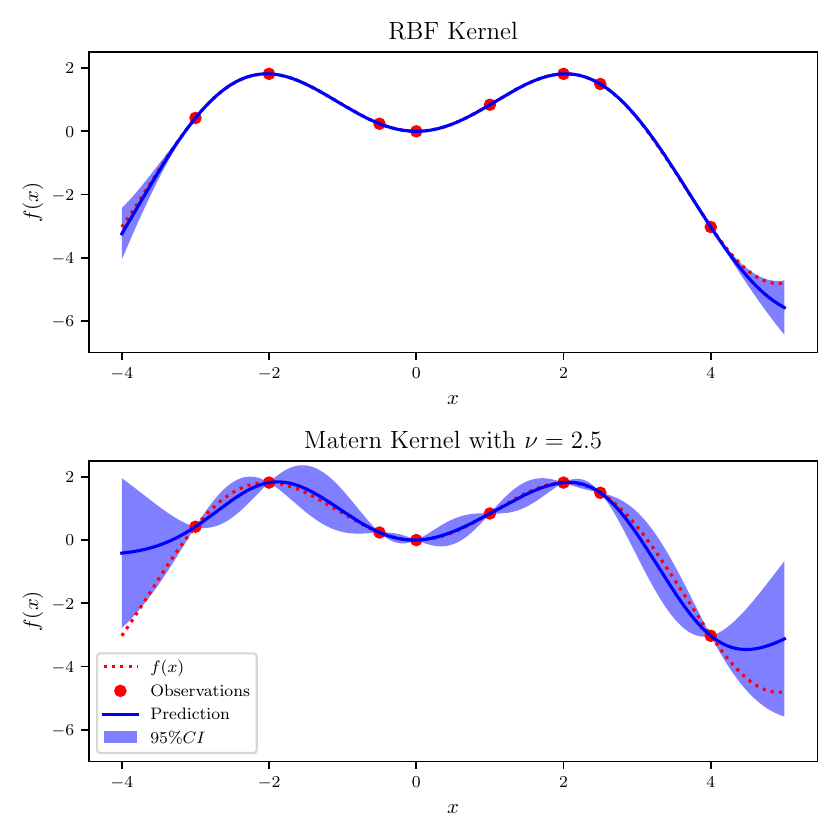}
    \caption{GP regression with RBF and Mat\'ern ($\nu = 5/2$) kernels}
    \label{fig:gp_1}
\end{figure}

Kernels functions play important roles in GP modeling. A GP model with the same parameter estimation method but different kernel functions may have different results. This can be seen in Figure \ref{fig:gp_1}. 

A simple function $f(x) = sin(x) * x$ is created and several points on $x$ are sampled as training data (red circles). As mentioned previously, the GP model yields both the mean and the standard deviation of the response, which allows us to plot the $95\%$ CI for each unobserved data point, see blue shaded area in Figure \ref{fig:gp_1}. Despite the prediction results (blue line) by two GP models showed good performance compared the the observations, the standard deviations yielded by two models are quite different. The RBF kernel showed small standard deviation while the model with Mat\'ern kernel has relatively large standard deviation. So extra caution should be paid when incorporating the error information of GP model to the total covariance $\bm \Sigma$ in Bayesian calibration. It is reasonably that different models have different confidence in their prediction, but the difference in the resulting total covariance matrix may lead to very different posterior distributions. So it is recommended that the uncertainty brought by GP surrogate should not be explicitly included in the total covariance. 

The GP model in this work is coded by the scikit-learn (\cite{sklearn_api}) open-source package in Python. The default parameter estimation method in scikit-learn is ``L-BGFS-B'' from scipy (\cite{2020SciPy-NMeth}). The parameter estimation aims at maximizing the log-marginal-likelihood (LML). The LML of GP may have multiple local optima, so the algorithm will restart automatically for a certain number of times to ensure an optimized set of hyper-parameter can be found. Cross Validation (CV) can also be used to estimate the hyper-parameters. Details about various parameter estimation in GP can be found in (\cite{rasmussen2004gaussian}). The accuracy of the GP emulator needs to be assessed before use. We are interested in the predictive accuracy at untried points, which can be done by quantifying the predictive error at an additional set of validation data (\cite{mc17-1, mc17-2}).

\subsection{Principal Component Analysis and Gaussian Processes}
When dealing with the time series regression problems, traditional regression methods becomes challenging as the highly-correlated output usually needs special treatment by many separate GP models which increases computational burden significantly. A multivariate GP surrogate model is required in this case instead of independent model for each output variable. For example, if there are a hundred measurements in a series, then all the hundred outputs need to be processed by the surrogate model, requiring significant higher training time. A dimension reduction of the high-dimensional output helps reduce the total computational burden and maintain the correlations of outputs. In this section, foundations of the PCA will be firstly introduced. A combined time series regression model by GP and PCA will be explained.

PCA can be viewed as a projection method which projects high-dimension data into a lower-dimension subspace such that the projected data are uncorrelated and have maximized explained variance (\cite{wu2018kriging}\cite{wangsurrogate}). PCA can be done through Eigen-decomposition of the covariance matrix or Singular Value Decomposition (SVD) of the data matrix. 

A brief mathematical introduction to PCA will be given here. Suppose our data matrix $\bm A$ has the shape $p \times N$, where $N$ stands for number of samples, $p$ stands for number of features (dimension of each output). This data matrix can be seen as the training dataset for the surrogate model. 

PCA is a linear transformation method where we transform the data matrix $\bm A$ by 
\begin{equation}
\label{eqa:pca1}
    \bm P \bm A = \bm B
\end{equation}
where $\bm P$ is a  transformation matrix and $\bm B$ is a  transformed data matrix. The goal of PCA is to find the proper $\bm P$ that makes rows of $\bm B$ uncorrelated and this can be done by SVD. If A is already centered (subtract row mean for each element), the SVD of A can be given by:
\begin{equation}
    \bm A = \bm U \bm \Lambda \bm V^T
\end{equation}
where $\bm U$ is an orthogonal matrix whose columns are the left-singular vectors of $\bm A$, $\bm V$ is an orthogonal matrix whose columns are the right-singular vectors of $bm A$, and $\bm \Lambda$ is a $p \times N$ diagonal matrix whose diagonal entries are the singular values of $\bm A$. The magnitudes of diagonal entries of $\bm \Lambda$ are then arranged in descending order and large singular values corresponds to important features in $\bm A$ (\cite{shlens2014tutorial}). Choose $\bm P$ as $\bm U^T$, so that
\begin{equation}
\label{eq:pca1}
    \bm P \bm A = \bm \Lambda \bm V^T = \bm B
\end{equation}

The covariance matrix of $\bm B$ is diagonal. The rows of $\bm P$ and the rows of $\bm B$ are called Principal Components (PC) and PC scores, respectively. 

The dimension reduction can then be performed by selecting a proper dimension of the principal subspace $p^*$, which is much smaller than $p$, but is able to make the principal components explain most of the variances. The first $p^*$ PCs form a $p^* \times p$ transformation matrix $\bm P^*$: 
\begin{equation}
    \bm P^* \bm A = \bm B^*
\end{equation}
where $\bm B^*$ is a $p^* \times N$ matrix whose rows represent new variables after dimension reduction. So, matrix $\bm B^*$ has much lower dimension than the original data matrix, and can be used as the training data for GP. 

\begin{figure}[!htbp]
    \centering
    \includegraphics[width =0.8\textwidth]{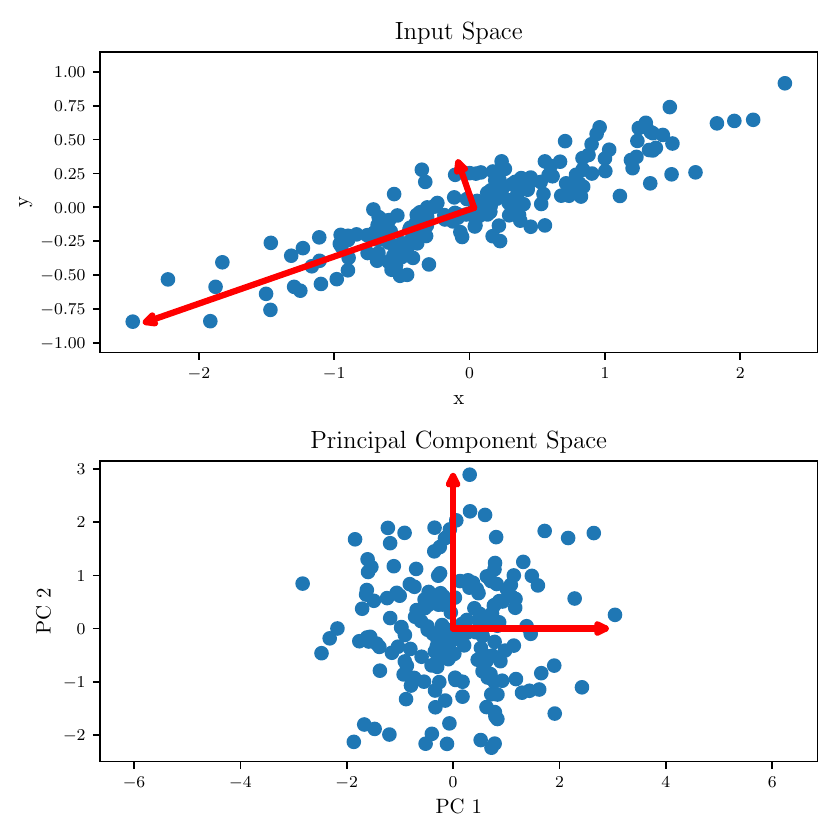}
    \caption{Illustration of PCA in 2D space}
    \label{fig:pca_1}
\end{figure}

The left matrix multiplication of the data matrix can be seen as rotating, moving, and scaling the coordinate system of the original data. Suppose we are looking at a 2 dimensional input shown in Figure \ref{fig:pca_1}, and we want to conduct dimension reduction on the data. It can be seen that the data has the potential to be reduced to one dimension because $x$ and $y$ are linearly related with some noise, knowing one of them can let us know roughly about the other. So the axes can be moved, rotated, and scaled as the red arrows show in Figure \ref{fig:pca_1}. The longest arrow (axis) can explain most of the variance of the data, while the data variation along the second arrow (axis) is small compared to the first.

\begin{figure}[!htbp]
    \centering
    \includegraphics[width =0.8\textwidth]{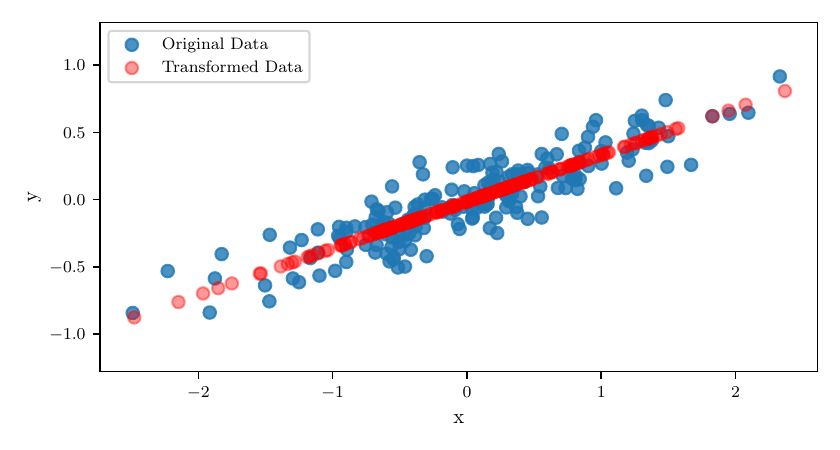}
    \caption{Comparison of original data and transformed data by PCA using one PC}
    \label{fig:pca_2}
\end{figure}

The transformed data $\bm B$ obtained by PCA can be seen in the lower plot of the Figure \ref{fig:pca_1}. The PCs are actually the projection of each data point onto the principal axes. From the upper figure  we can see that PC1 explains most of the variation so we select PC1 only and transform it back to the original input space, the results are shown in Figure \ref{fig:pca_2}. We can see the linear relationship between $x$ and $y$ is still captured after dropping PC2, and the transformed data using only 1 principal component is ``good enough'' to represent the original data.

From Figure \ref{fig:pca_2}, we can have a better understanding of that PCA does for dimension reduction: the information along the least important axes is removed, and the component(s) of data which explains most of the variance is(are) left. The fraction of the variance that is cut out can be seen as a measure of how much information is discarded by the dimension reduction of PCA. This metric is usually used to determine the number of PCs to select.

\begin{figure}[!htbp]
    \centering
    \includegraphics[width =0.7\textwidth]{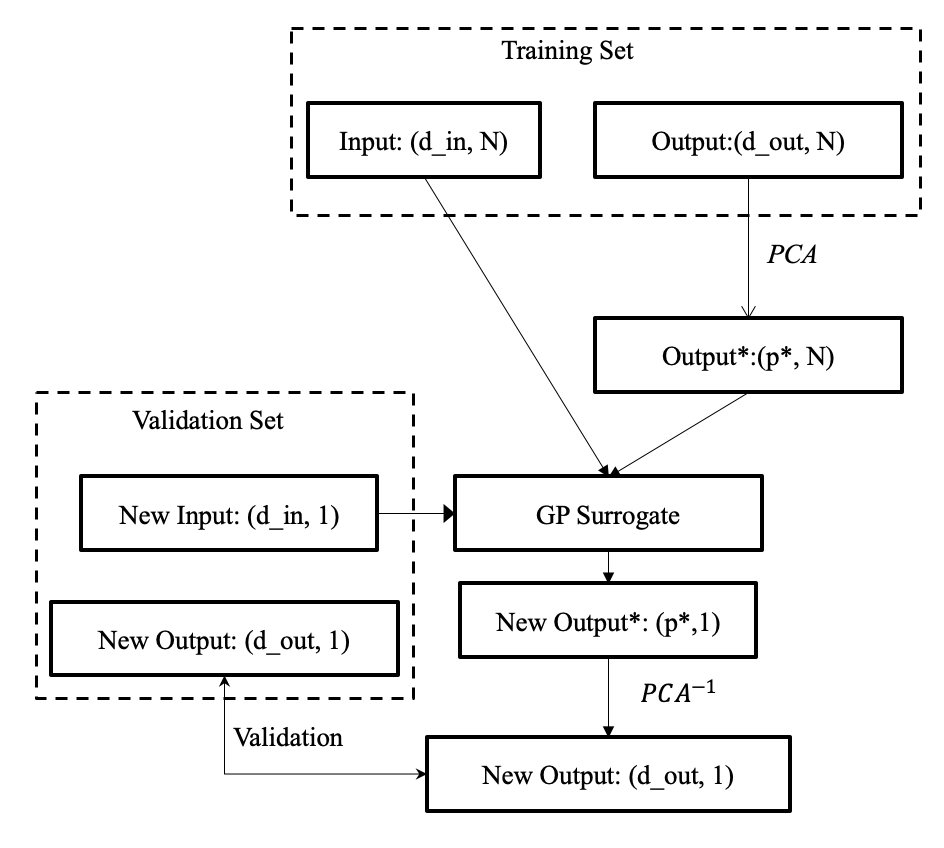}
    \caption{Flowchart of the GP + PCA regression model for high-dimension and high-correlation outputs}
    \label{fig:gppca_1}
\end{figure}

In the reduced dimension, we would be able to combine GP model with PCA to construct a regression model. The flowchart of using GP and PCA as regression model for the high-dimension and high-correlation problem is shown in Figure \ref{fig:gppca_1}.

The dataset should be split into a training set and validation set so that over-fitting can be avoided and the model performance can be objectively quantified. In the training set with $N$ samples, suppose the the input and the output has the dimension d\_in and d\_out, respectively. The output we are dealing here is time-dependent data so it has high dimensions and high correlation. Applying PCA to the output and selecting proper number of principal components $p^*$, we can reduce the output dimension to $p^*$. The selection of number of PCs should be based on the fraction of variance explained by the PCs. A GP model can then be built here based on the original input (d\_in, N) and the output ($p^*$, N) in the reduced dimension. 

In regression, we are interested in predicting the output at an new untried point, and the prediction by GP model here gives us the output in the reduced dimension with shape ($p^*, 1$). Inverse of PCA can then be applied here to transform the data into the original output space. The inverse can be done simply by left multiplication of $\bm P^{-1}$ to Equation \ref{eq:pca1}. If the predicted output is compared with the new output in the validation set, the performance of the overall model can then be quantified.

In summary, the combined GP and PCA model requires building one PCA model and $p^*$ GP models, which reduces the computational burden of building d\_out GP models. The dimension reduction becomes possible because these time-dependent outputs are highly correlated.

\subsection{Artificial Neural Network}
Artificial Neural Network (ANN) is a powerful tool for regression tasks for its capability to model complex patterns in datasets using multiple hidden layers and non-linear activation functions. In many cases, Neural Network may seem like an overkill as regression models since it is mostly known for its capability of handling very complex and computationally expensive jobs such as image processing, natural language process, artificial intelligence, etc. However, NN or even Deep NN is nothing but repeated combinations of simple linear and non-linear transformations. Their structures can be designed in a very sophisticated way but the smallest unit in DNN is still a perceptron. So ANN models can be seen as convenient and powerful regression models if properly tuned. 

Perceptron (or neuron) is the smallest unit in a neural network. The output of a perceptron is simply an activation function applied to linear combination of inputs. Say if the input $\bm x$ has $n$ dimensions, now the output from a perceptron if $f(b + x_1w_1+ x_2w_2 + ...+ x_nw_n)$, where $b, w_i$ are weights and $f(.)$ is the activation function. Many activation functions are available and they have different characteristics to be used in different scenarios. Most commonly used activation function for regression purposes include linear or Rectified Linear Units (ReLU) (\cite{ramachandran2017searching}). Linear activation provides simple unbounded output and is easy to solve, but is limited in its complexity and have less power in complex problems. ReLU is widely used in many ANN structures and it ranges from 0 to infinity. It has the form:
\begin{equation}
    f(x) = \begin{cases}
    0,  x < 0 \\
    x,  x \geq 0
    \end{cases}
\end{equation}
which gives an output itself if positive and 0 otherwise. It is a good approximator and any function can be approximated with a combination of Relu~\cite{ramachandran2017searching}.

Mathematically, a one layer network has the form:
\begin{equation}
    \bm y = \bm w^T \bm x
\end{equation}
and a two layer network has the form:
\begin{equation}
    \bm y = \bm w^T \sigma_1(\bm A_1 \bm x + \bm b_1)
\end{equation}
where $\bm A$ is the weight matrix and $\bm b$ is the intercept vector of that layer. Iterating the process, a fully connected network with $L$ layers has the form:
\begin{equation}
\label{eqa:ann1}
    \bm y = \bm w^T \sigma_1(\bm A_1 \sigma_2(...\bm A_{L-2}\sigma_{L-1}(\bm A_{L-1} \bm x + \bm b_{L-1}) + \bm b_{L-2} ...) + \bm b_1)
\end{equation}

The objective function of NN model in this regression task can be defined as the sum of squared errors: $\Sigma (\hat y_i - y_{true})^2$, which is usually referred to as L2 norm loss function. The parameters in the model ($\bm A, \bm b$) can then be estimated by minimizing the objective function with respective to $\bm A$ and $\bm b$. Taking derivatives of equation \ref{fig:ann_1} with respect to these weight parameters is straightforward and can be scaled to any finite depth because of the chain rule. The method of adjusting weight parameters to minimize the objective function using the chain rule is called backpropagation (\cite{hecht1992theory}). With the development of computer science and computational power, the backpropagation algorithm can be applied to very large-scale neural networks and has achieved successes in many fields.

ANN model should be used with special caution in regression tasks because it is very prone to overfitting due to its complex structure. To avoid overfitting, there are generally two methods to considered when the number of available data points can not be increased:
\begin{itemize}
    \item Weight Regularization. Add a penalty term to the error function during the training process, so that weight parameters can be constrained. Because large weighs tend to cause sharp transitions in the activation functions and thus large changes in output for small changes in inputs (\cite{reed1999neural}).
    \item Add dropout layers. During the training process, we randomly drop units (along with their connections) from the neural network so that it prevents units from adapting too much with each other  (\cite{srivastava2014dropout}). 
\end{itemize}

Both two methods have been widely used to prevent overfitting. One additional benefit of adding dropout layers is that model uncertainty can also be calculated besides the point estimate (\cite{gal2016dropout}), which gives us confidence interval for the prediction. So dropout method is preferred in many application where uncertainty information is needed such as reinforcement learning and Bayesian optimization (\cite{snoek2012practical}).

In this work, there are two main motivations of using ANN as surrogate models. The first is that we need to deal with high dimensional output in the time-series data, and ANN is capable of dealing with high dimensions with its flexible structure. The second is that some efficient MCMC algorithms require derivatives of the model prediction, which is easy to obtain from Equation \ref{eqa:ann1}. The ANN model is coded using the TensorFlow package (\cite{tensorflow2015-whitepaper}) in Python.

\subsection{Construction and Validation of Surrogate Model}

\subsubsection{GP and PCA Model}

A surrogate model needs to be built for each transient. In this section, we take 6T-flow reduction transient as an example to illustrate how the surrogate model is built and validated. In previous section, we have explained in detail the process of building the GP + PCA model for the time series regression model. Now we will go through the process using the 6T-FR transient as an example.

\begin{figure}[!htbp]
    \centering
    \includegraphics[width = 0.6\textwidth]{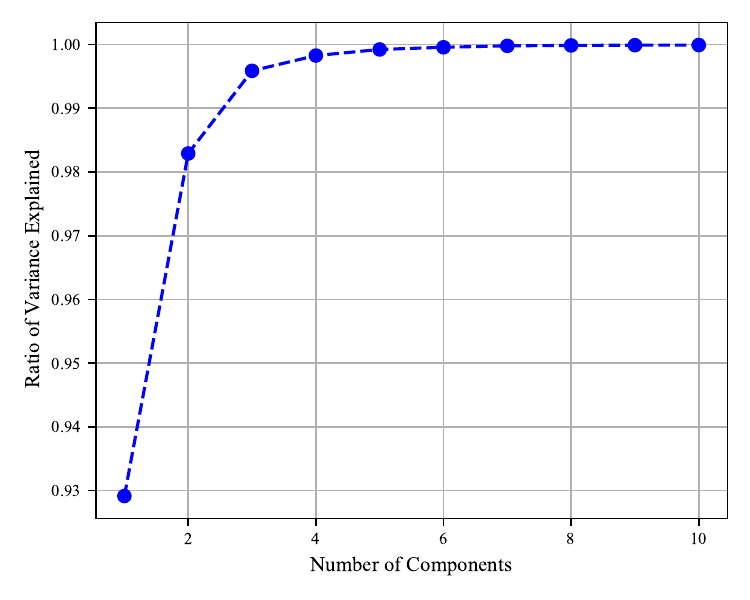}
    \caption{Ratio of explained variance v.s. number of principal components}
    \label{fig:6gppcavalie2}
\end{figure}

In the 6T-FR transient, there are 40 time steps and 3 outputs, so the output dimension is $3 \times 40 = 120$. Now we firstly use PCA as a dimension reduction tool to reduce the original dimension (120) to a much smaller number, and this number is just the number of principal components we will select. 400 samples are firstly drawn using LHS in the range of $\theta \in (0,5)$ as the training set. PCA is then conducted on this $400\times 120$ matrix and the ratio of explained variance versus the number of principal components is shown in Figure \ref{fig:6gppcavalie2}. We can see that 4 principle components can lead to a very good 0.999 percentage of the variance, which means that the selected 4 PCs are a good enough representation of the original 120 dimensional data.

\begin{figure}[!htbp]
    \centering
    \includegraphics[width = 0.6\textwidth]{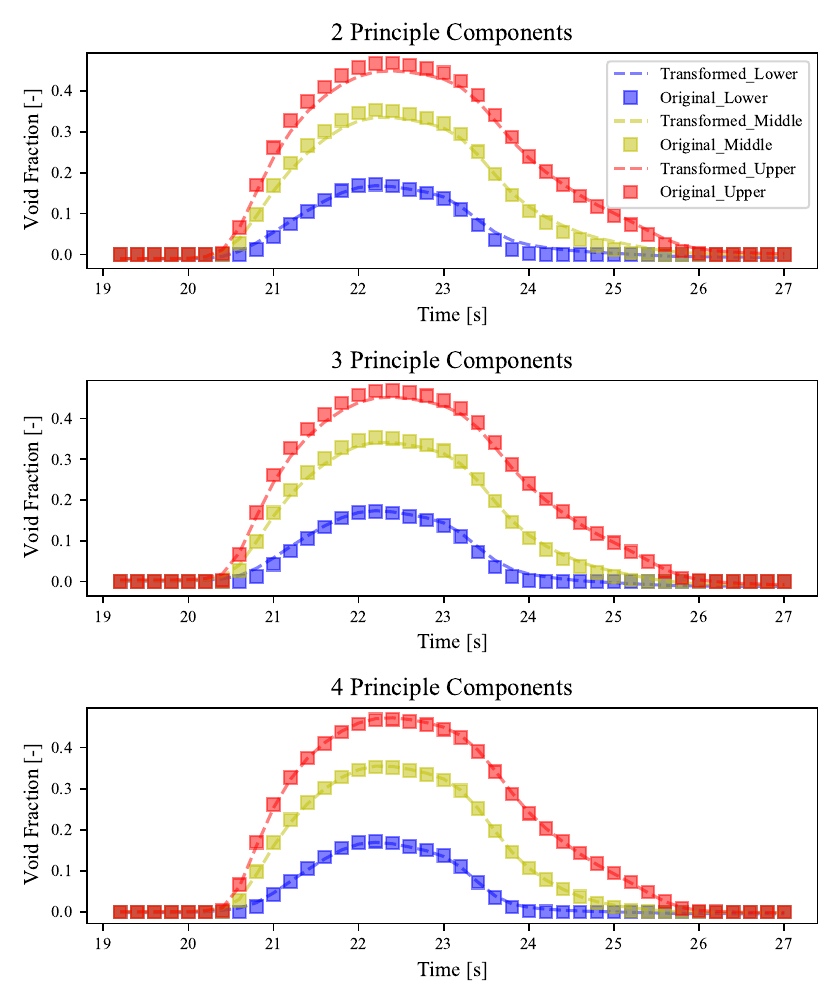}
    \caption{Comparison between original data (TRACE) and back-transformed data using 2, 3, and 4 PCs. The discrepancy between the curves are decreasing as more PCs are used.}
    \label{fig:6gppcavalie3}
\end{figure}

To further analyze the effect of the number of PCs, we can transform the reduced data back to the original space, and see if there is any discrepancy between the transformed data and the original data. Mathematically, this is equivalent to comparing $\bm P^{-1} \bm B^*$ with original data matrix $\bm A$. (Recall that in PCA, $\bm A = \bm P^{-1}B$, $\bm P^*A = \bm B^*$ and $\bm B^*$ is the matrix after dimension reduction. See equation \ref{eqa:pca1}.) The comparison results for one of the samples are shown in Figure \ref{fig:6gppcavalie3}, where 2, 3, and 4 PCs are tested. In the upper plot of Figure \ref{fig:6gppcavalie3}, we can see the red squares are slightly different than the dotted lines, indicating a loss of accuracy using 2 PCs. The results with 3PCs show better agreement but slight difference can still be observed. When 4 PCs are used, we can see that the transformed data and the original date agree almost perfectly. 

The same analysis needs to be conducted for the other 8 transients because they have different profiles and may need more PCs to achieve good accuracy. Results will not be repeated here for simplicity. Finally, we found out that 5 PCs are enough for very good accuracy for all transients, and we will use 5 PCs as the output to train GP regression models. We have successfully reduced the output dimension from 120 to 5 using PCA thanks to the correlations in the time-dependent data.

\begin{figure}[!htbp]
    \centering
    \includegraphics[width = 0.6\textwidth]{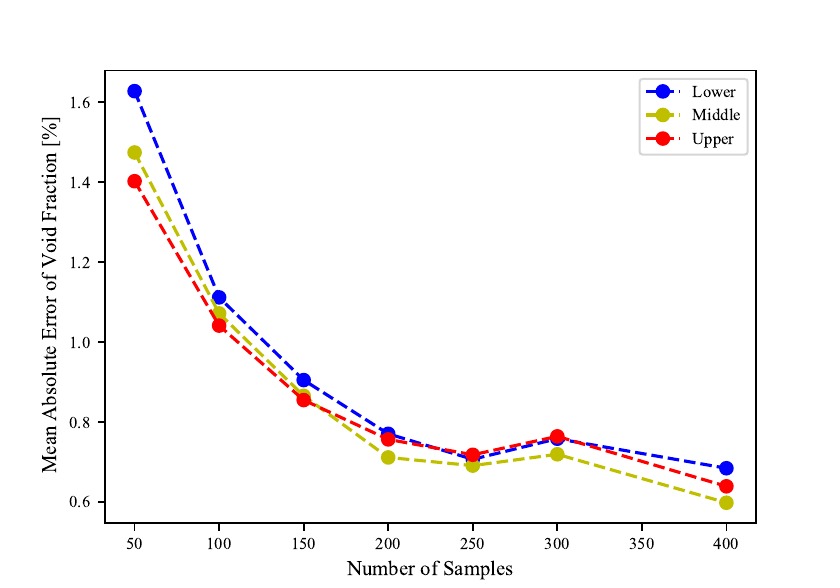}
    \caption{Convergence study of GP PCA model}
    \label{fig:6vali}
\end{figure}

Using the 5 PCs, we can train GP models whose inputs are physical models parameters and outputs are these PCs. Now we can quantify the accuracy of the combined GP + PCA model as a whole. 50 randomly selected samples are used as a testing set. The error is calculated as the difference between original TRACE simulation results and the predicted results by GP + PCA model, on the testing set. Note that each sample corresponds to a time series so the error is calculated for multiple time points in a series. The mean absolute error versus the number of training samples is plotted in Figure \ref{fig:6vali}. We can see that after 200 samples, the errors are below 1\% and the curve remains relatively flat. But 400 samples lead to the lowest MAE so we will use 400 samples for this study. This may seem computationally expensive to draw so many samples from TRACE, however, since this is a one-time expense and the transients simulations don't take a very long time so it is acceptable to sacrifice several hours to achieve better accuracy.

\subsection{Artificial Neural Network Model}
Although GP+PCA model showed very good predictive performance on the high-dimension high-correlation output, it has two disadvantages: (1) the model structure is complex and not friendly to users; (2) it is not easy to calculate the derivative of the model responses with respective to inputs, which is required in NUTS algorithm for efficient posterior sampling. In this work, we developed an ANN model, which is capable of handling high-dimension output, has a straightforward model structure, and provides convenient derivative information. 

The downside of ANN is that its model structure is so flexible that it requires a lot of tuning work in order to find an optimal structure for the given task. The current problem requires a 4-dimensional input and 120-dimensional output (for 7T-FR transient only, output shape may be different for other transients). So we employ a fully connected NN with 4 nodes in the first(input) layer and 120 nodes in the last(output) layer and use linear activation function in the output layer since it is a regression problem. The structure of the hidden layers are determined by several trials and errors, the structure that leads to the best overall prediction performance in the validation set is selected. The final model has two hidden layers with 32 nodes in each, so the final NN model is connected by $4 \times 32 \times 32 \times 120$ nodes. L2 norm is applied to reduce the overfitting issue of the NN model. 

\begin{figure}[!htbp]
    \centering
    \includegraphics[width = \textwidth]{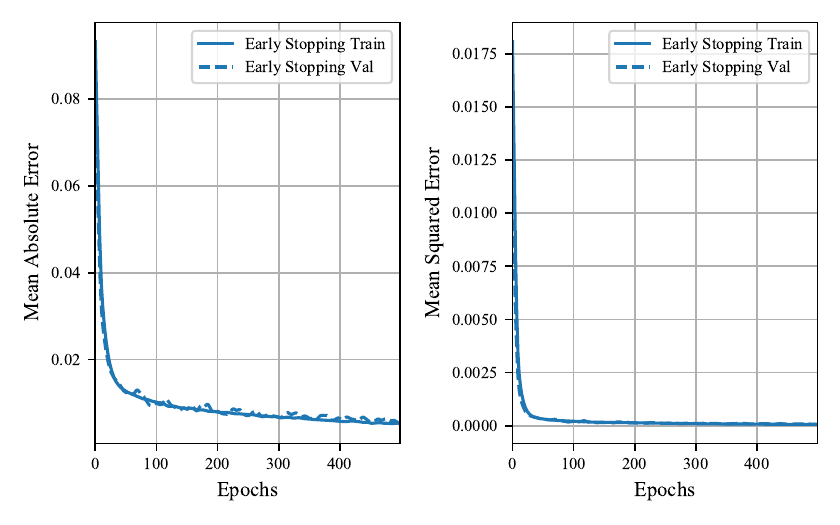}
    \caption{Training history of ANN}
    \label{fig:annvali1}
\end{figure}

The model is trained on the same 400 samples used for GP+PCA in the previous section, extra 50 random samples are used as the validation set. Figure \ref{fig:annvali1} shows the training history of the ANN model, and we can see that both regression MAE and MSE decreased to a low level after a certain number of iterations. The error in the validation set (dotted lines) also converged and overfitting didn't occur. 

\begin{figure}[!htbp]
    \centering
    \includegraphics[width = 0.7\textwidth]{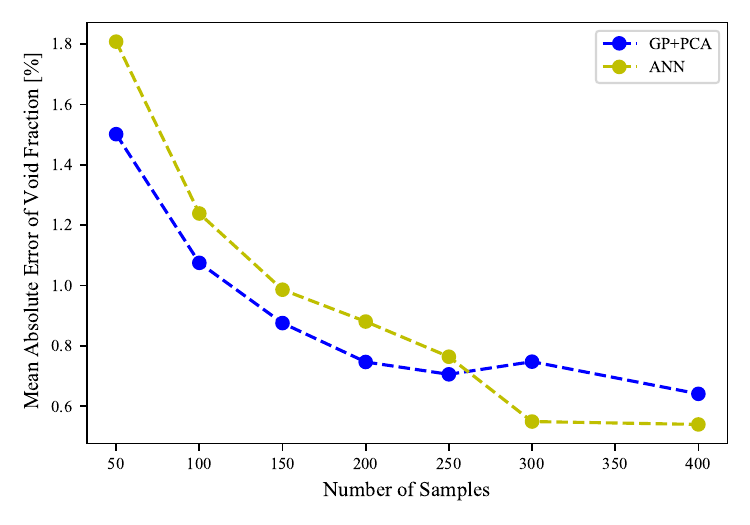}
    \caption{Convergence study for GP+PCA and ANN}
    \label{fig:annvali2}
\end{figure}

The performance of the ANN model and GP+PCA model are compared in the convergence study shown in Figure \ref{fig:annvali2}. MAEs of void fraction predictions are compared using different number of samples for training. We can see that the GP+PCA model outperforms the ANN model when there are less than 250 samples, while the ANN model showed better predictive capability with more data. 

\section{Results of Hierarchical Bayesian Model for Time-Dependent Problems}
\label{sec5}

\subsection{Covariance in Time-Dependent Data}
Now that we have solved the issue regarding surrogate models for time-dependent data, let's go back to the model updating equation to review some assumptions. In the model updating equation when model discrepancy is not considered, the experimental observations actually follow multivariate normal distribution $\bm X \sim \mathcal{N}(\bm \mu, \bm \Sigma), \bm X \in \mathcal{R}^k $, where $\bm \mu$ can be considered as the true noise-free value provided by simulation model, and $\bm \Sigma$ is the covariance matrix, which contains uncertainty information of experiment and code.  Then the probability density function of $\bm X$ follows:
$$
f(\bm x) = (2\pi)^{-\frac{k}{2}}\det(\bm \Sigma)^{-\frac{1}{2}}\exp(\frac{1}{2}(\bm x - \bm \mu)^T \bm \Sigma^{-1} (\bm x - \bm \mu) )
$$

Traditionally, we have been treating $\bm \Sigma$ as a diagonal matrix, meaning that we assumed $\bm X$ is independently normally distributed and ignored $cov(X_i, X_j)$ when $i \neq j$. This is a valid assumption in most steady-state experiments, since measurements were usually conducted independently. However, in transient cases, there can be high correlation between the measurement errors at different time steps, causing $\bm \Sigma$ to be non-diagonal. The correlated measurement error has been studied in many fields such as hydrologic models by \cite{tiedeman2013effect}), Meteorology models by \cite{stewart2008correlated}, etc. There can be various reasons for the error correlation, such as spatially and temporally correlated anomalies in the underlying physical phenomena, multiple observations derived from a single measurement, etc (\cite{tiedeman2013effect}). The correlated observation error in the transient experiments in the PSBT benchmark can be observed from Figure \ref{fig:ts_5vali}, \ref{fig:ts_6vali}, and \ref{fig:ts_7vali}. For example, in the flow reduction transient in Figure \ref{fig:ts_7vali}, the boundary conditions are stable after around 163$s$, so the void fractions measured at three locations are expected to be also stable (normally distributed at certain levels). However, we can see clear serial correlation in those time series, indicating correlated errors.

The off-diagonal element in covariance matrix is calculated by $\Sigma_{i,j} = E[(X_i - \mu_i)(X_j - \mu_j)]$, which requires a bunch of samples of the measured time series data. The samples cannot be obtained realistically since experiments cannot be repeated, but we can ``simulate'' the experiment process and the error by our simulation model TRACE, using the uncertainty information from boundary conditions. Since the measured uncertainty in boundary conditions is known to us, we can propagate its uncertainty through the code and treat the resulting model responses' uncertainty as a proxy of the real measurement uncertainty in time series. Since the code is deterministic we also need to add independently distributed error to it during the calibration process. It should be noted that this method can be seen as an over-estimation of the covariance matrix, because the correlation would exist in the whole time series using this method. However, from what we can observe in Figure \ref{fig:ts_7vali}, the correlations only exist in multiple time steps and not the whole time series. There are also other methods available for estimating the observation covariance matrix, but none are without fault (\cite{waller2016theoretical}\cite{stewart2013data}).

\begin{figure}[!htbp]
    \centering
    \includegraphics[width = 0.7\textwidth]{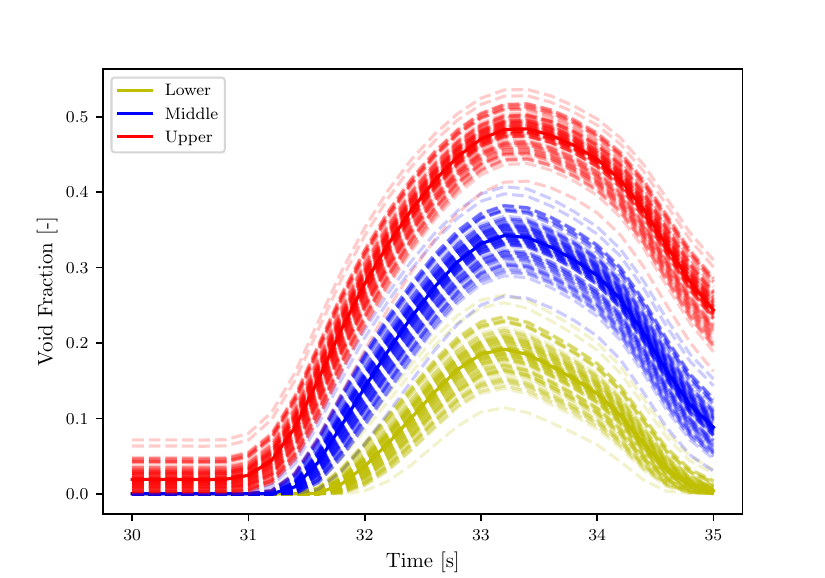}
    \caption{Simulated uncertainty in 5T-Flow Reduction transient}
    \label{fig:uq-bc1}
\end{figure}

Figure \ref{fig:uq-bc1} shows the uncertainty of void fraction caused by boundary conditions uncertainties in 5T-Flow Reduction transient. Using this ensemble of time-series realizations, we can calculate the covariance matrix to be used in the posterior distribution.
 
\begin{figure}[!htbp]
    \centering
    \includegraphics[width = 0.7\textwidth]{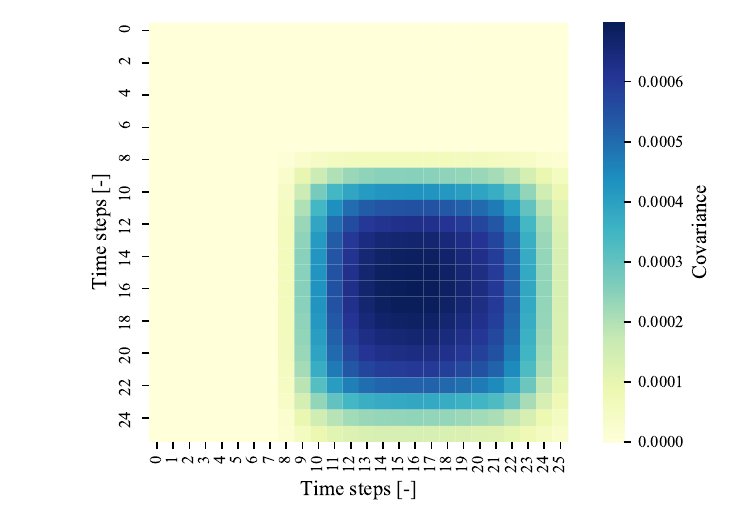}
    \caption{Heatmap of covariance matrix using data from the Lower measurement location in Figure \ref{fig:uq-bc1}.}
    \label{fig:cov_heat}
\end{figure}

The heatmap of the covariance matrix calculated by samples of the 'Lower' output (yellow curves) in Figure \ref{fig:uq-bc1} is shown in Figure \ref{fig:cov_heat}. We can see that the correlations between time steps do exist and we should take that into consideration. Same procedure are applied to the other 8 transients and covariance matrix 

\subsection{Posterior Sampling Results and Validation}

Now we can use the obtained covariance matrix in the formulation of posterior, and sample posterior distributions using the MCMC algorithm. In this section, we will demonstrate the calibration results using a single transient 7T-Flow Reduction, and compare the results when covariance is included and not included.

\begin{figure}[!htbp]
    \centering
    \includegraphics[width = 0.8\textwidth]{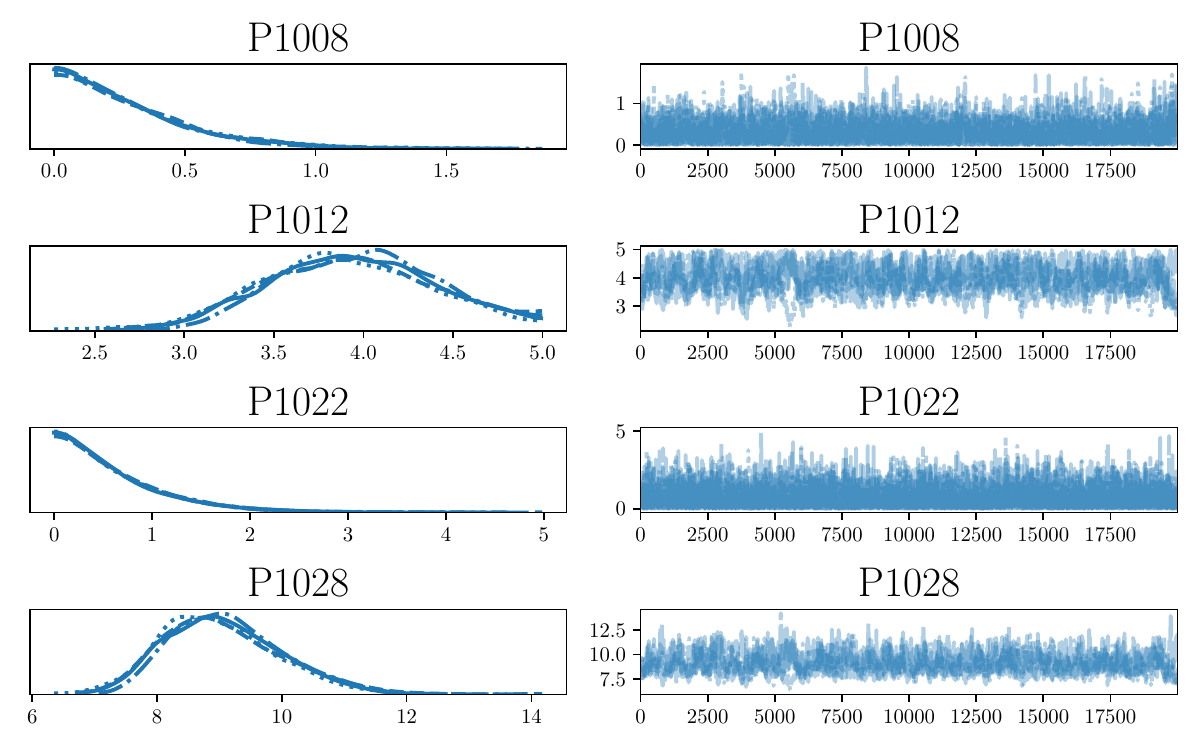}
    \caption{Posterior distribution and trace plots calibrated by 7T-FR, covariance not considered}
    \label{fig:7tft_nocov_1}
\end{figure}

\begin{figure}[!htbp]
    \centering
    \includegraphics[width = 0.8\textwidth]{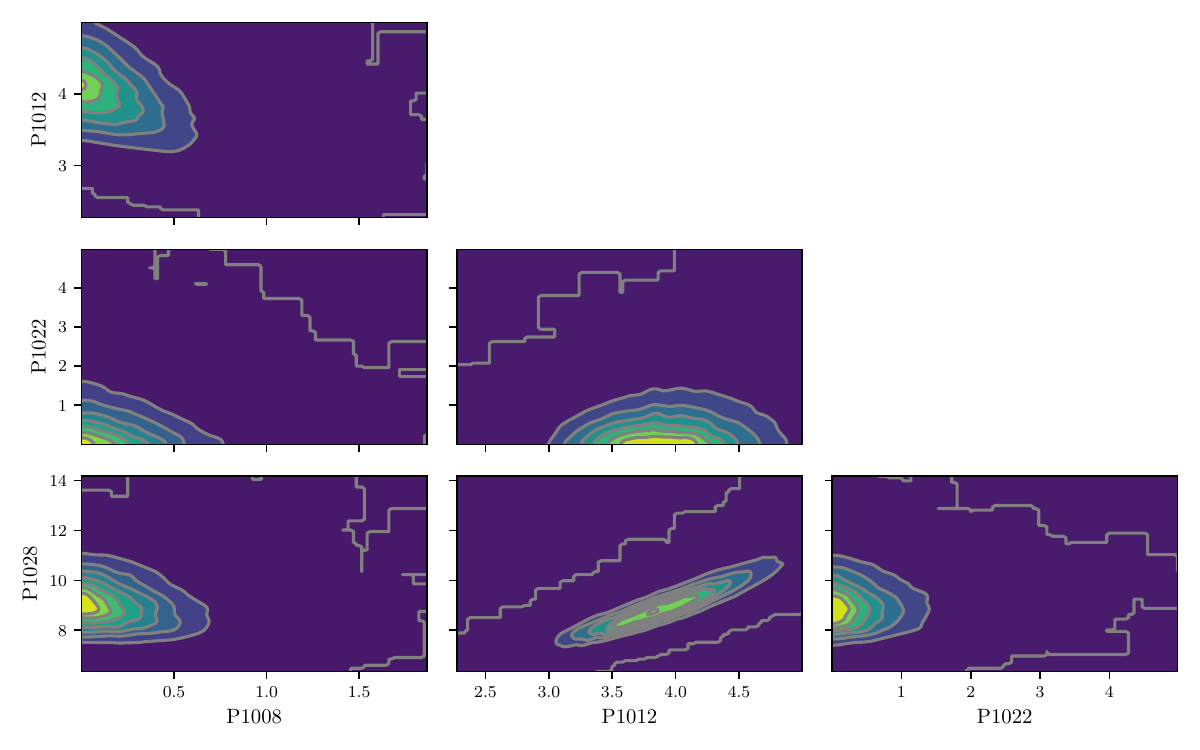}
    \caption{Pair-wise joint distribution calibrated by 7T-FR, covariance not considered}
    \label{fig:7tft_nocov_2}
\end{figure}

Because of the usage of the ANN model, the efficient NUTS algorithm can be used for posterior sampling. We firstly show results when covariance is not considered. In this case, the procedure is exactly the same as calibrating steady-state data. In this transient experiment, 151 data points are available, so we can simply calibrate the physical model parameters against these data, with the covariance matrix being diagonal. Figure \ref{fig:7tft_nocov_1} shows posterior distributions and trace plots of the 4 physical model parameters, and Figure \ref{fig:7tft_nocov_2} shows the pair-wise joint distribution among the four parameters. We can see that the chain is converged and mixes well. Two parameters `P1008' and `P1022' tend to 0 which can be an indication of model discrepancy. `P1028' the wall drag coefficient also deviates far away from its nominal values. Here the range of `P1028' is larger than the original range of (0,5) because the iterative re-sampling procedure is used to ensure the posterior can be fully included.  Correlation of input parameters exist between `P1008' and `P1012', and between `P1012' and `1028', as can be seen in Figure \ref{fig:7tft_nocov_2}.

\begin{figure}[!htbp]
    \centering
    \includegraphics[width = 0.8\textwidth]{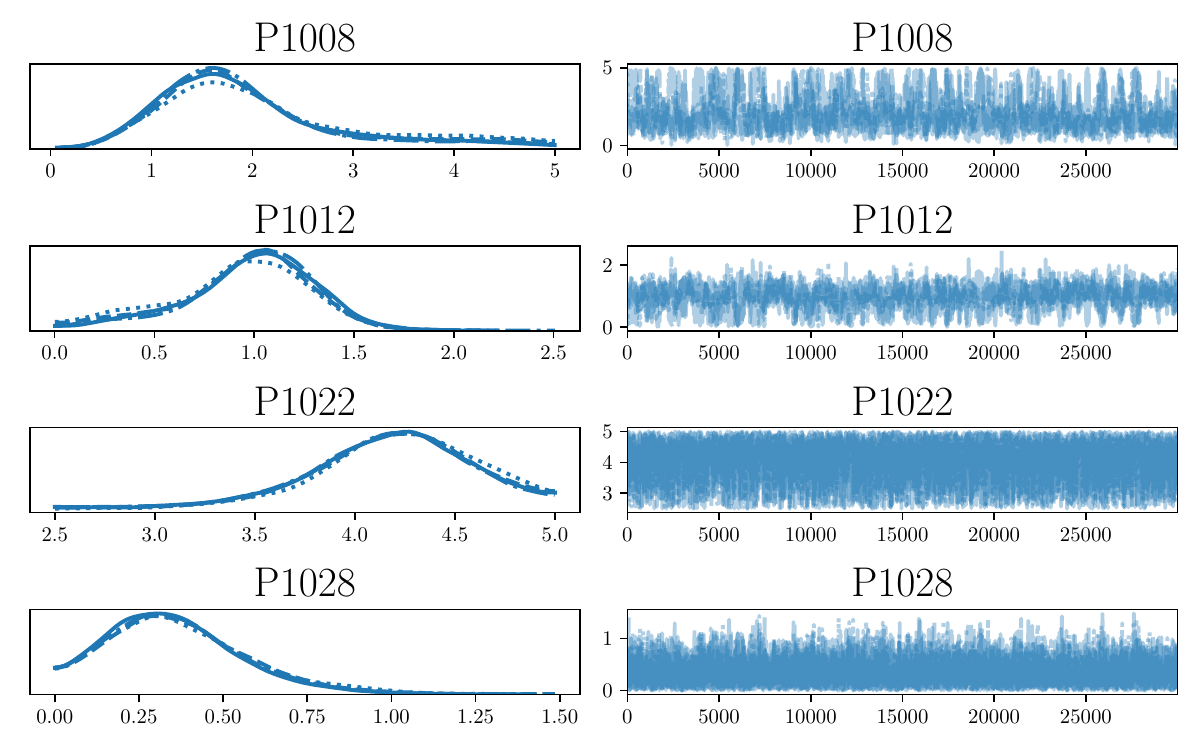}
    \caption{Posterior distribution and trace plots calibrated by 7T-FR, covariance is considered}
    \label{fig:7tft_cov_1}
\end{figure}

\begin{figure}[!htbp]
    \centering
    \includegraphics[width = 0.8\textwidth]{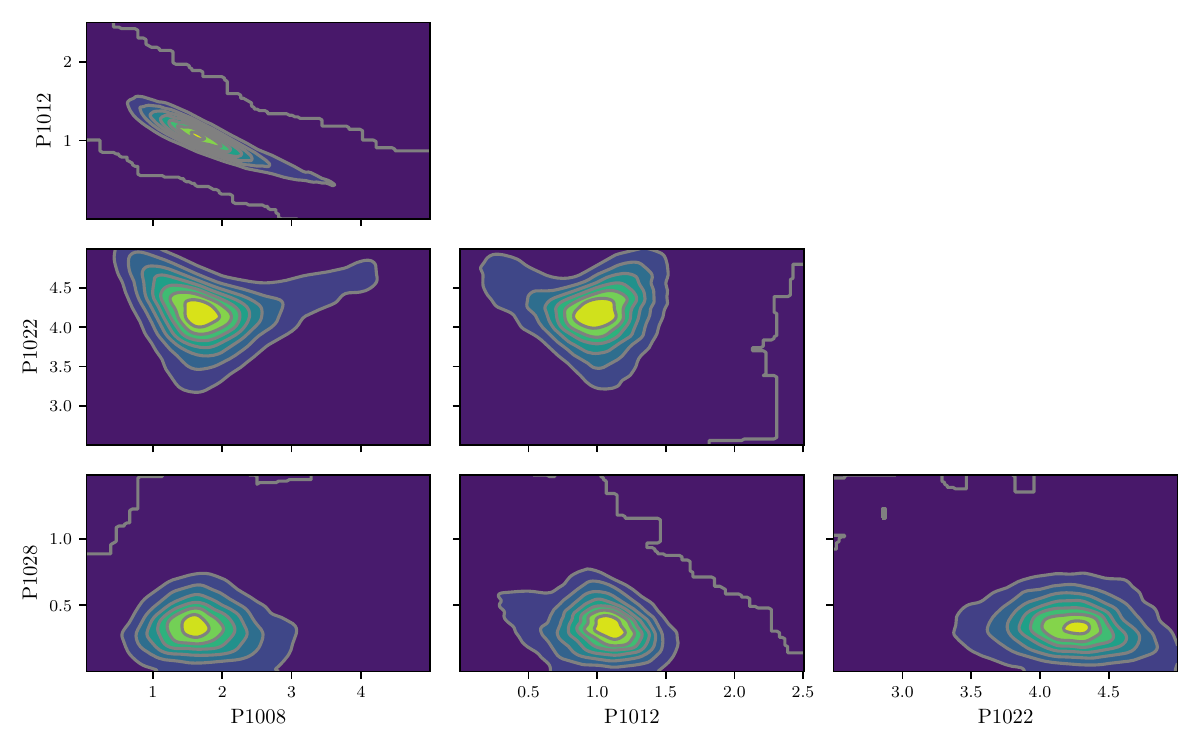}
    \caption{Pair-wise joint distribution calibrated by 7T-FR, covariance is considered}
    \label{fig:7tft_cov_2}
\end{figure}

Now we include the covariance matrix in the posterior sampling, the posterior distributions, trace plots and pair-wise joint distributions are shown in Figure \ref{fig:7tft_cov_1} and \ref{fig:7tft_cov_2}. We can see the results are very different from Figure \ref{fig:7tft_nocov_1} and \ref{fig:7tft_nocov_2}, where covariance is not considered. Now significant input correlation only exist between 'P1008' and `P1012', which is the same as we have observed in steady-state cases in the previous section.

\begin{figure}[!htbp]
    \centering
    \includegraphics[width = \textwidth]{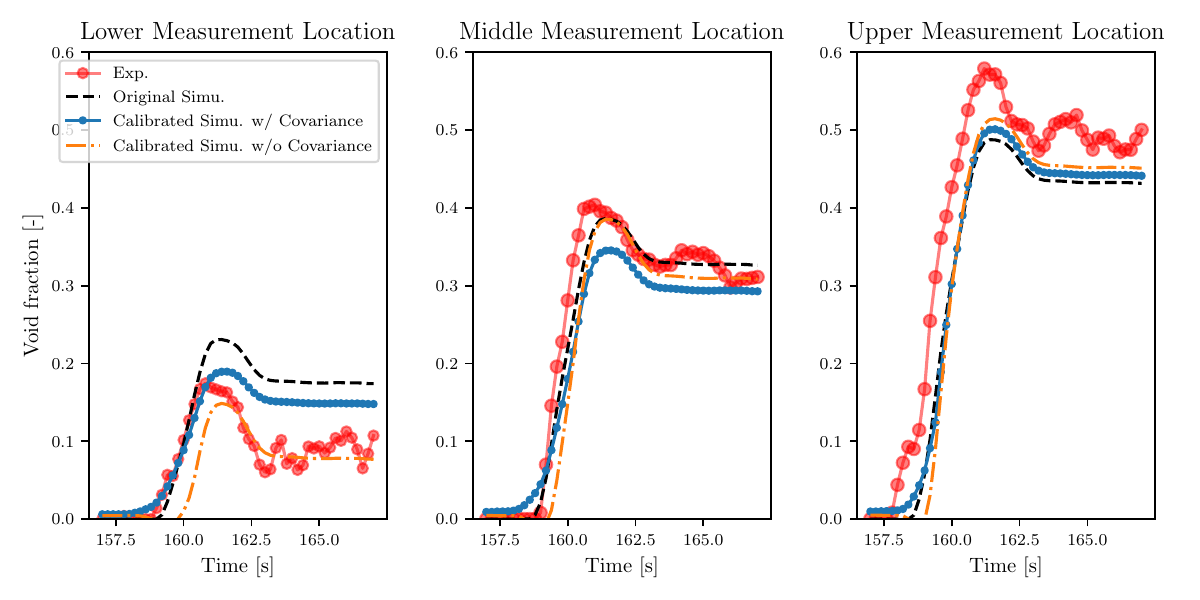}
    \caption{Comparison of model responses with different calibration parameters}
    \label{fig:cov_vali}
\end{figure}

In order to determine which result is more reasonable and should be used, in Figure \ref{fig:cov_vali} we compare the model responses using different physical model parameters: (1) $\theta_{prior} = 1$ (black dotted line), (2) $\theta_{post1}$ when covariance is considered (blue line with circles), and (3) $\theta_{post2}$ when covariance is not considered (orange dotted lines). Experiment data (red circles) is also included for comparison.

Comparing the results with and without covariance in Figure \ref{fig:cov_vali}, we can see that the result with covariance information tends to be closer to the experiment during transition periods, when covariance is small (see Figure \ref{fig:cov_heat}). While the result without covariance information simply finds the curve that may lead to the smallest overall error level. For example in the left figure, the curve without covariance chooses to ``fit'' the tail part of the time series, but fails to follow the transition period when void fraction is increasing at around 160s. So now we can better understand the role of the covariance matrix in Bayesian calibration of time series data: it encourages us to put more ``weight'' on the overall shape of the time series rather than focusing on reducing averaged error. 

\subsection{Using the Hierarchical Framework for Time-Dependent Problems}
\label{sec:6-hb}
We have demonstrated the effects of covariance information on the results of calibration through one transient case. However, as we see in Figure \ref{fig:cov_vali}, the large model discrepancy may lead to significant over-fitting issue: the posterior distribution of $\bm \theta$ obtained from one transient would not help achieve better agreement between simulation and experiment for other transient cases. In light of this situation, we propose to use the hierarchical model to avoid over-fitting.

Similar to the hierarchical structure shown before, we assume that calibration parameter $\bm \theta$ can be different for each group. In this scenario, each group represents one transient case, and there are 9 transients in all, we use 5 of them as the training (calibration) set and the other 4 of them as the testing set. Five transients randomly selected for training are: `5T-FR',`5T-PI',`6T-FR',`6T-TI',`7T-FR'.
The priors of parameters are defined below:
\begin{align*}
    \mu \_P1008 & \sim \mathbf{Unif}(0,5)\\
    \sigma \_P1008 & \sim \mathbf{Unif}(0,1)\\
    P1008 & \sim \mathcal{N}(\mu \_ P1008,\sigma \_P1008)
    %\sigma & \sim \mathbf{Unif}(0,1)
\end{align*}

\begin{figure}[!htbp]
    \centering
    \includegraphics[width = 0.9\textwidth]{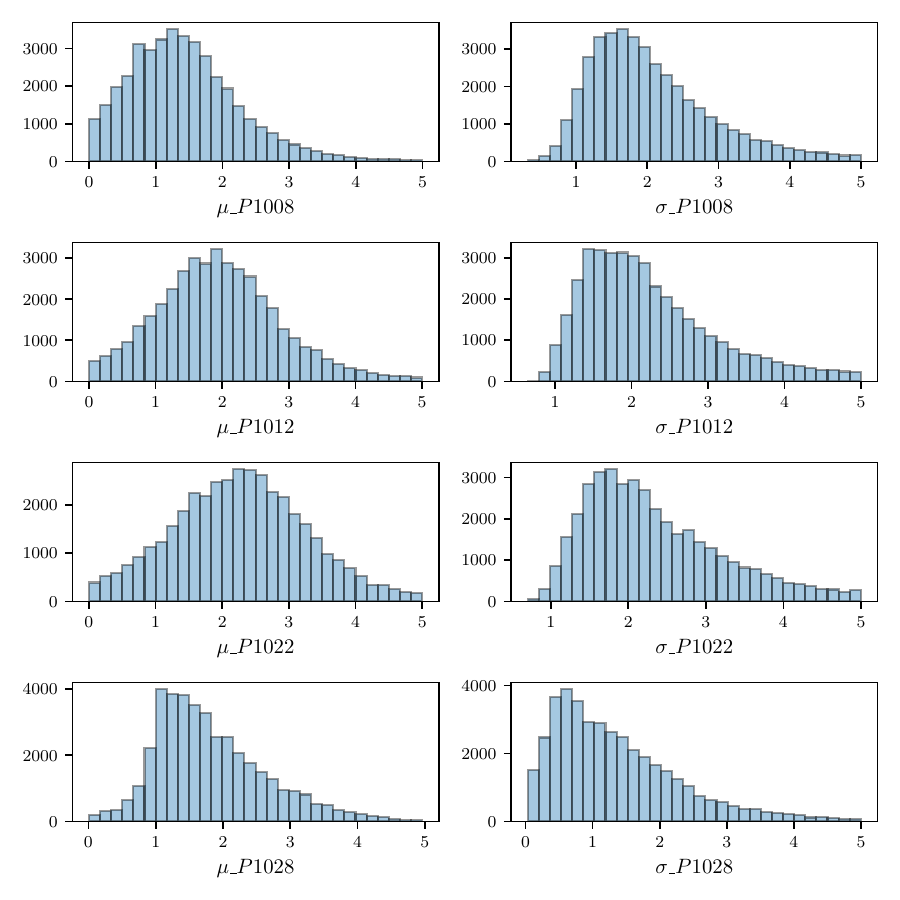}
    \caption{Posterior distributions of shared parameters in the hierarchical model}
    \label{fig:post_hb_ts}
\end{figure}

\begin{figure}[!htbp]
    \centering
    \includegraphics[width = 0.9\textwidth]{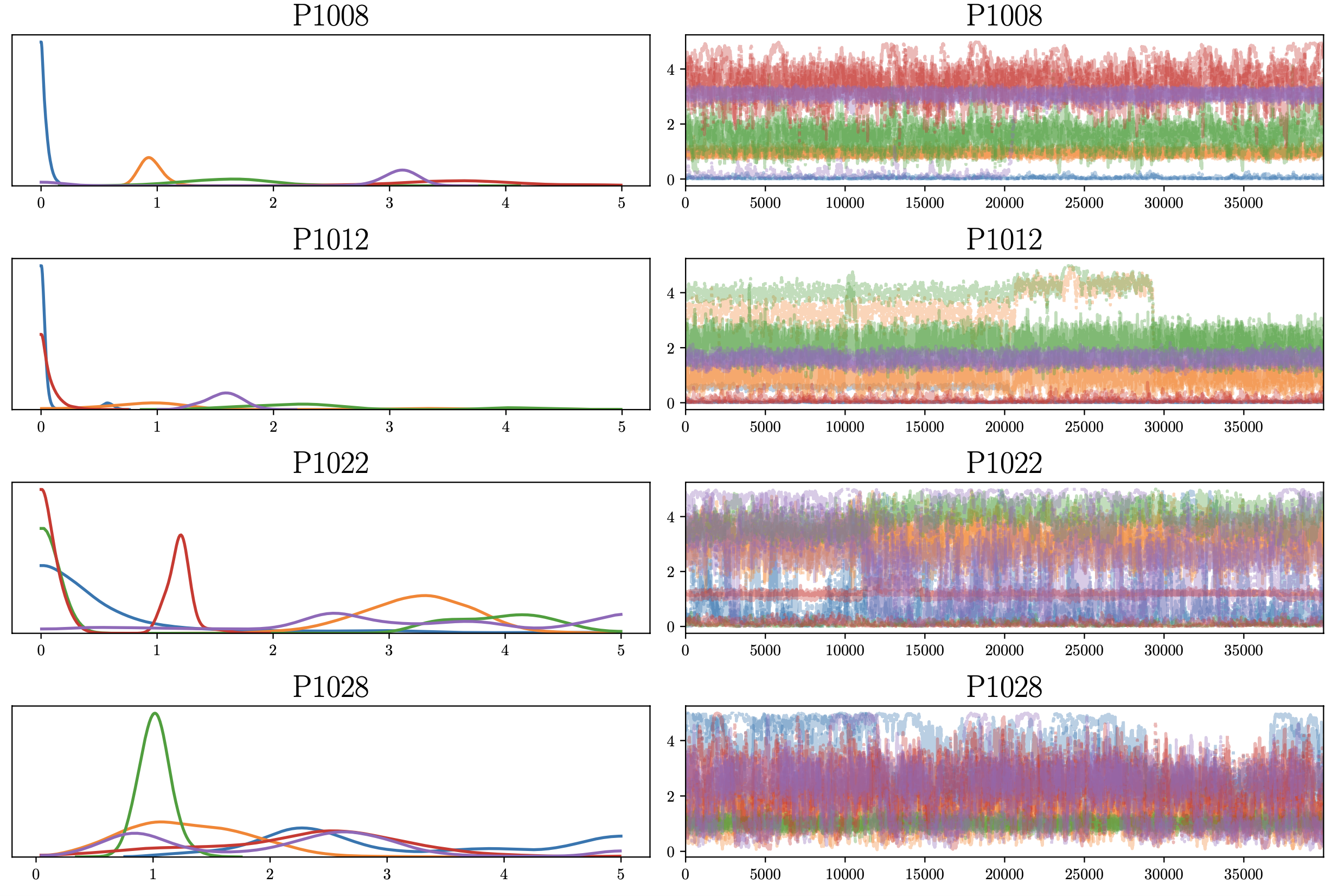}
    \caption{Posterior distributions of per-group parameters in the hierarchical model}
    \label{fig:post_hb_ts_ind}
\end{figure}

Using the NUTS method we are able to obtain converged posterior samples for all the hyperparameters (shared parameter) and per-group parameters. Figure \ref{fig:post_hb_ts} shows the posterior distribution for all hyperparameters, and Figure \ref{fig:post_hb_ts_ind} shows the posterior for each per-group parameter and their trace plots. The cluster-specific parameter have relatively large differences, as we can see in Figure \ref{fig:post_hb_ts_ind}. This is possibly due to the large model discrepancy in some cases, some parameter are overfitting the model discrepancy instead of showing their own distributions. The hierarchical model actually takes the overfitting issue into consideration by allowing these parameters to be in a common distribution, and we can take this distribution as our updated knowledge about those parameters, which is a conservative measure and can avoid overfitting. For example, for parameter `P1008', we can use $\mathcal{N}( \overline{{\mu\_P1008}}, \overline{{\sigma\_P1008}})$ as the posterior, where $\overline{{\mu\_P1008}}$ and $\overline{{\sigma\_P1008}})$ can be obtained by their posterior samples.

\begin{figure}[!htbp]
    \centering
    \includegraphics[width = 0.9\textwidth]{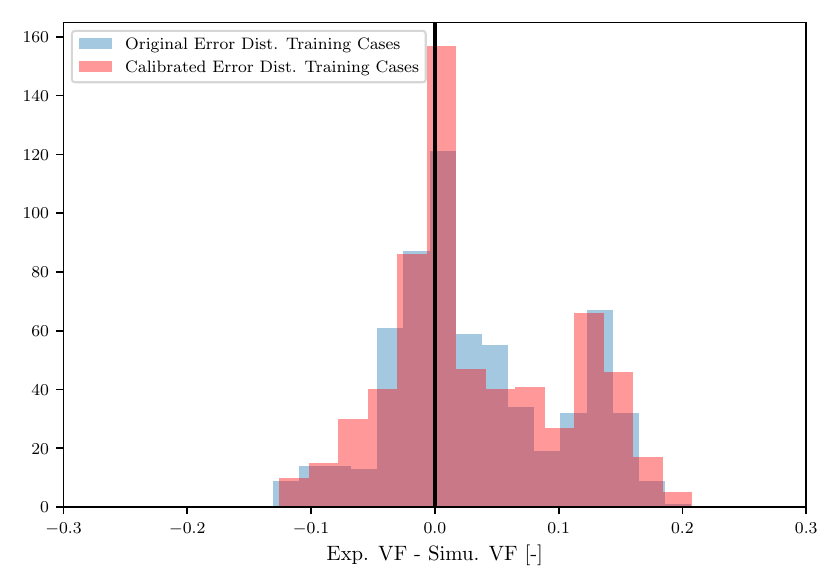}
    \caption{Comparison of error distributions in the training set }
    \label{fig:hb_ts_vali1}
\end{figure}

\begin{figure}[htbp]
    \centering
    \includegraphics[width = 0.9\textwidth]{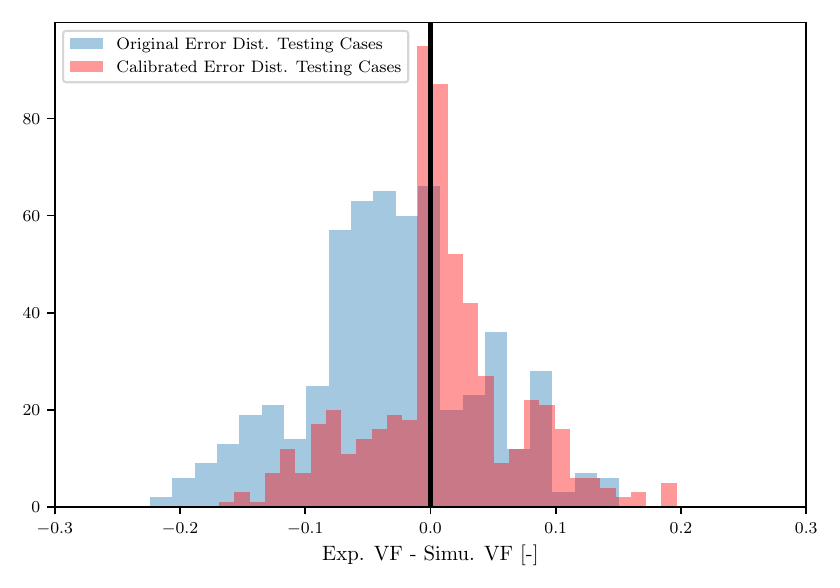}
    \caption{Comparison of error distributions in the validation set}
    \label{fig:hb_ts_vali2}
\end{figure}

Now we can validate the posteriors using the validation set, the validation method is the same as we have been using in previous sections. Since multiple time series are more difficult to visualize, we directly compare the distribution of errors (Exp. void fraction $-$ Simu. void fraction) using calibrated values off the calibration parameter $\bm \theta$. The comparison results for training set and for the validation set are shown in Figure \ref{fig:hb_ts_vali1} and \ref{fig:hb_ts_vali2}, respectively. We can see that the error distribution is slightly improved in the training set, and is significantly improved in the testing set. The fact that the calibrated model performance in the validation set is better than in the training set demonstrates the capability to avoid overfitting in hierarchical models.

\section{Summary}
\label{sec6}

This work tackles the IUQ problem for time-dependent datasets in nuclear TH applications. A key advancement of this study was the incorporation of covariance information in posterior distribution calculations, demonstrating how this inclusion better aligns the simulation results with experimental data, especially during transition phases of the transients. This method emphasizes capturing the time series' overall shape rather than simply minimizing average errors.

The central role of surrogate models in Bayesian calibration is underscored, highlighting their utility in simplifying the computationally intensive simulations. Artificial Neural Networks proved particularly effective for handling high-dimensional outputs and for facilitating derivative calculations within the MCMC algorithms, outperforming other models when applied to certain TH applications.

The research also identified the risk of over-fitting with traditional IUQ approaches in individual transients. To mitigate this, a hierarchical model was utilized and validated, which showed improved error distributions for both training and validation sets. This confirms the hierarchical model's superiority in avoiding over-fitting, thus presenting a robust solution for IUQ in time-dependent nuclear TH simulations.


\bibliographystyle{elsarticle-harv} 
\bibliography{main}



%%\end{thebibliography}
\end{document}